\documentclass[aps,pra,twocolumn,balance,superscriptaddress,showpacs,preprintnumbers,amsmath,amssymb,widetext,,floats,a4paper]{revtex4}
\usepackage{latexsym}
\usepackage{dcolumn}
\usepackage{amsmath}
\usepackage{epsf}
\usepackage{float}
\usepackage{hyperref}
\usepackage{gensymb}

\usepackage{graphicx}
\usepackage{dcolumn}
\usepackage{bm}

\usepackage{color,ulem}
\newcommand{\be}{\begin{equation}}
\newcommand{\ee}{\end{equation}}
\newcommand{\bea}{\begin{eqnarray}}
\newcommand{\eea}{\end{eqnarray}}

\begin{document}

\title{Lepton-neutron  interaction and S-wave low energy parameters}

\author{Jaume Carbonell}
\affiliation{Université Paris-Saclay, CNRS/IN2P3, IJCLab, 91405, Orsay, France}

\author{Tobias Frederico}
\affiliation{Instituto Tecnol\'ogico de Aeron\'autica, 12.228-900 S\~ao Jos\'e dos Campos, Brazil}

\date{\today}

\begin{abstract}
A lepton-neutron potential in configuration space is obtained.
It is based on the  Coulomb plus hyperfine interaction Hamiltonian integrated over the neutron charge and magnetic densities.
Different parametrisations of the neutron electromagnetic form factors are compared.
It is given in the operator form with a central, spin-spin, tensor and spin-orbit terms.
The potentials for lowest partial waves states are presented.
We compute the lepton-neutron lepton ($ln$)  low-energy parameters for the S-waves,  estimate
the zero-energy cross sections for higher angular momentum states, and point out a possible divergence in the partial wave summation due to the  spin-orbit potential.
\end{abstract}

\maketitle


\section{Introduction}

The lepton-neutron ($ln$) interaction is dominated by electromagnetic effects.
At the leading order, they are due to the electric interaction between the point-like lepton ($l$) and the neutron ($n$) internal charge distribution,
 to the magnetic interaction between  $l$  and $n$ magnetic moments and to the  coupling between the $n$ magnetic
 moment in the field created by the $l$ current.
They may have different relative signs and strengths depending on the lepton flavour as well as on the quantum number of the  $ln$ system
and,  despite of its perturbative character, offer a rich variety of non trivial behaviours.
A key point in their theoretical estimation is to properly take into account the neutron's internal electromagnetic structure, 
obtained through the corresponding  electric ($G^n_E$) and magnetic ($G_M^n$) form factors.

The lepton-neutron low-energy parameters (LEP) are fundamental quantities which {are} worth to 
estimate and measure.
Furthermore, they  might have several applications  in the  precision atomic spectroscopy measurements   {using $e$'s and $\mu$'s \cite{Nancy_PRL126_2021},
 in determining the deuteron 
\cite{mud_CREMA_Science353_2016} and $\alpha$-particle
charge  radius \cite{mu_4He_Nature589_2021},
as well as in solid state physics with low energy $n$ scattering on materials \cite{Sears_PR141_1986,Koester_PRC51_1995,Abele_PPNP_60_2008}.
Future experiments based on muonic X-ray spectroscopy are also proposed to significantly improve the charge radii of light nuclei \cite{QUARTET_Physiscs6_2024}
as well as some beyond the standard model investigations related to, still speculative, new bosons (see e.g. \cite{New_Bosons_Arxiv24}).}	

The aim of the present article
is to {    obtain a $ln$ potential  in configuration space} allowing  us to compute, within a non-relativistic dynamics,
the LEP parameters as well as the corresponding phase shifts and cross sections for the lowest partial  waves.
It is based on the Hyperfine Hamiltonian integrated over the $n$ charge and magnetic densities.
The potential has four components: a central part due to Coulomb interaction, a spin-spin and a tensor term due to the  dipole-dipole magnetic interaction,
and spin-orbit term coupling the $ln$ relative  angular momentum, $L$, to the $n$ spin $s_n$.
This potential is the keystone to evaluate the electronic effects in the low energy neutron scattering in nuclear atomic targets.

Section~\ref{Sect_rhons} is devoted to describe some selected $n$ electromagnetic form factors  
 used to derive the corresponding charge and magnetic densities in configuration space.

The $ln$ electromagnetic potential in configuration space is  obtained in section~\ref{Sect_Vln} 
and the main properties of this interaction in the lowest partial wave are discussed.
 
The numerical results for the  $ln$ low-energy scattering observables are summarised in Section~\ref{Sect_Results}, 
with special emphasis in the  (S-wave) low-energy parameters, phase shifts and zero-energy cross sections  and the scattering of $n$ with electrons-bound-to-atoms 
(Sub. \ref{SSect_S}) and a subsection devoted to the zero-energy scattering with higher partial waves (Sub. \ref{SSect_P}). 
Some final remarks conclude this work in section ~\ref{Sect_Conclusion}.

\section{Neutron densities} \label{Sect_rhons}

The $n$ -- charge   $\rho^n_{c}(\vec{r})$  and magnetization   $\rho^{n}_m(\vec{r})$ -- densities  can be obtained
by Fourier transforming the corresponding  Sachs electric ($G_E$) and magnetic ($G_M$) form factors in the Breit frame~\cite{Lorce_PRL125_2020}:
\begin{multline}\label{rho_n}
\rho_{c,m}^n(\vec{r})  =\int  \frac{d\vec{q}}{(2\pi)^3}  \;  G^n_{E,M}(q^2) \; e^{i \vec{q} \cdot \vec{r}} \\
\quad\Longleftrightarrow\qquad  G^n_{E,M}(q^2)= \int d\vec{r} \; \rho_{c,m}(\vec{r}) \;  e^{ i \vec{q} \cdot \vec{r} } \, , 
\end{multline}
where $t=q^2=-Q^2$ is the space-like momentum transfer. By expanding the plane wave in the right-hand side of (\ref{rho_n})
\[ e^{   i  \vec{q} \cdot \vec{r}  }   =   1+  i  (\vec{q} \cdot \vec{r})    -   {1\over 2} (\vec{q} \cdot \vec{r})^2   
-   {i\over 6} (\vec{q} \cdot \vec{r})^3+  {1\over 24} (\vec{q} \cdot \vec{r})^4 + \ldots \] 
and integrating over the angular part, one obtains
\begin{equation}\label{Gq2_exp}
 G(q^2)= G(0) -{\langle r^2\rangle\over 6}   q^2   + {\langle r^4\rangle\over 120}   q^4 + O(q^6)  \, . 
\end{equation}
The (even) radial moments $\langle r^{2k}_n\rangle_{c,m}$  of the $n$  charge and magnetic distribution can be alternatively obtained as  $k$-derivatives of the corresponding form factors
with respect to $q^2$:
\begin{equation}\label{r2k}
   \langle r^{2k}_n\rangle_{c,m} =   \;  {(-1)^k \,k! \over (2k+1)!} \;  \left[{d^kG_{E,M}\over d(q^2)^k}\right]_{q^2=0} \, , 
\end{equation}
with $\, k=1,2,\ldots$.

\subsection{Charge density}

The $n$-charge density $\rho_c^n$ satisfies
\[ \int d\vec{r} \;  \rho_c^n(\vec r)=0\, , \]
and must reproduces the experimental value of the $n$ mean squared  charge radius \cite{PDG_2022}:
\[ \langle r^2_n\rangle= \int d\vec{r} \; r^2 \rho_c^n(\vec r)=        {-}\;   0.116 \pm 0.002 \;   { {\rm fm}^2} \]

If we assume for $G^n_E$ the simple phenomenological form,  suggested by {    Friar~\cite{FN_ANP8_1975}}
\begin{equation}\label{GEn_F}
G_E^n(q^2) =\beta_n {q^2\over \left( 1+ {q^2\over b_n^2} \right)^3} \, ,
\end{equation}
with parameters $b_n$=4.27 fm$^{-1}$ and $\beta_n$=0.0189 fm$^2$, one gets
\begin{equation}\label{rhocn_F}
\rho_c^n(\vec{r})     = {(\beta_nb_n^2)b_n^3\over32\pi} \; (3-x) \; e^{-x} \qquad r=b_n\;x\, ,
\end{equation}
and a $n$ charge mean squared radius $\langle r^2_n\rangle= -0.113 \;{\rm fm}^2$   {(in elementary charge units $e$)}.
Despite its simplicity, Friar form factor~(\ref{GEn_F}) gives quite accurate results and allows simple analytical expressions.
It was also used in~\cite{AV18_WSS_PRC51_1995} for computing the electromagnetic corrections to the nucleon-nucleon (NN) S-wave low-energy parameters
and in~\cite{KVM_PRC69_2003}   to compute  $n$-deuteron scattering observables.

\begin{figure}[htbp]
\centering\includegraphics[width=8.5cm]{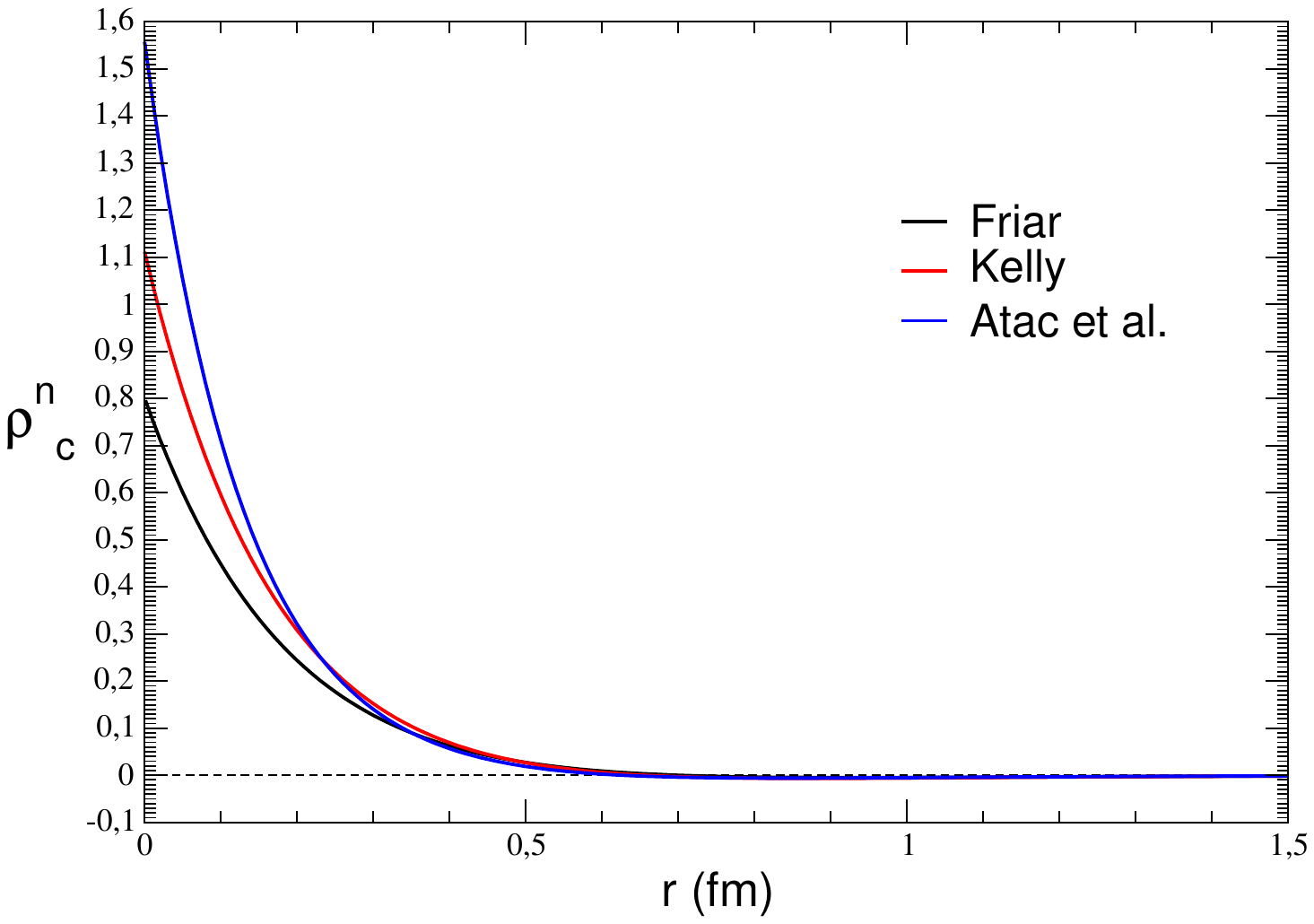}
\centering\includegraphics[width=8.5cm]{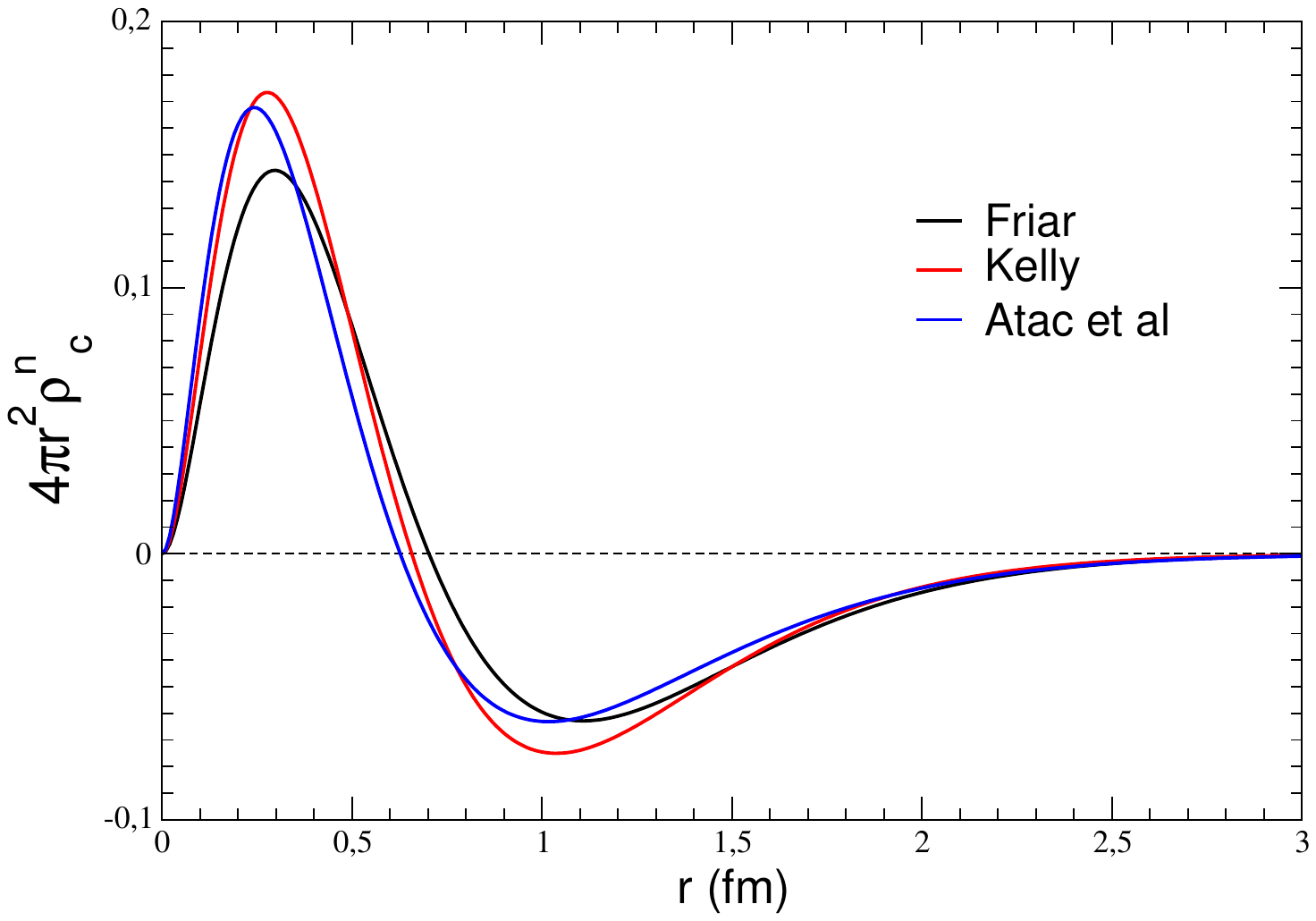}
\caption{Neutron charge densities   {(in elementary charge units $e$)}: $\rho_c^n(r)$ (upper panel)  and $4\pi r^2\rho_c^n(r)$ (lower panel)   {obtained
with different $n$ charge form factors: Friar (\ref{GEn_F}), Kelly  \ref{GEn_Kelly})  and Atac et al. \cite{Atac_Nature_Com_2021}}.}\label{Fig_rho_n}
\end{figure}

For the sake of completeness, we have  also considered  the  more accurate $n$ charge densities proposed by {    Kelly~\cite{Kelly_PRC70_2004}}.
It has  the form
\begin{equation}\label{GEn_Kelly}
G_E^n(q^2)=  {A\tau\over 1+ B\tau} G_D(q^2)    \qquad \tau={q^2\over 4m_p^2} \, ,
\end{equation} 
where 
\begin{equation}\label{G_Dipole}
 G_D(q^2)= {1\over \left( 1 + {q^2\over b^2} \right)^2}  
 \end{equation}
 is the dipole form, $b=4.27 \;{\rm fm}^{-1} \,[ b^2=  0.71 (GeV/c)^2 ]$ and the dimensionless parameters $A=1.70\pm0.04$, $B=3.30\pm0.32$ were adjusted
to reproduce the experimental data.\par

The corresponding charge density is
\begin{multline}\label{rhocn_Kelly}
 \rho^n_c(\vec{r}) = A'\;  {(b\beta)^2 \over 8 (\beta^2-b^2)^2 \pi}   \;{b^5\over x}\;\\ \times \left\{   \left[x + {\beta^2\over b^2} (2-x)  \right] e^{-x}  - 2 {\beta^2\over b^2}  e^{-\beta r} \, ,  \right\}  
\end{multline}
with $A'={A\over 4m_p^2}$=   0.01879769  fm$^2$ and $\beta= {2m_p\over \sqrt B}$=5.234983  fm$^{-1}$.
It gives a  $n$ charge radius $\langle r^2_n\rangle=-0.112\pm0.003$\,   {fm$^2$}.

A new parametrisation of the Kelly form factor was recently  proposed by {    Atac et al.~\cite{Atac_Nature_Com_2021}}
with the values $A=1.655\pm0.126$, $B=0.909\pm0.583$. This gives $\langle r^2_n\rangle=-0.110$\;   {fm$^2$}.

The corresponding $n$-charge densities are represented in  Fig.~\ref{Fig_rho_n}.
Despite reproducing well the experimental $n$ charge radius, they lead to sizeable different results at small values of $r$ (factor 2) as well as a $20\%$ 
difference in the zero of $\rho^n_c(r)$. 

\subsection{Magnetic density}

\begin{figure}[htbp]
\centering\includegraphics[width=8.cm]{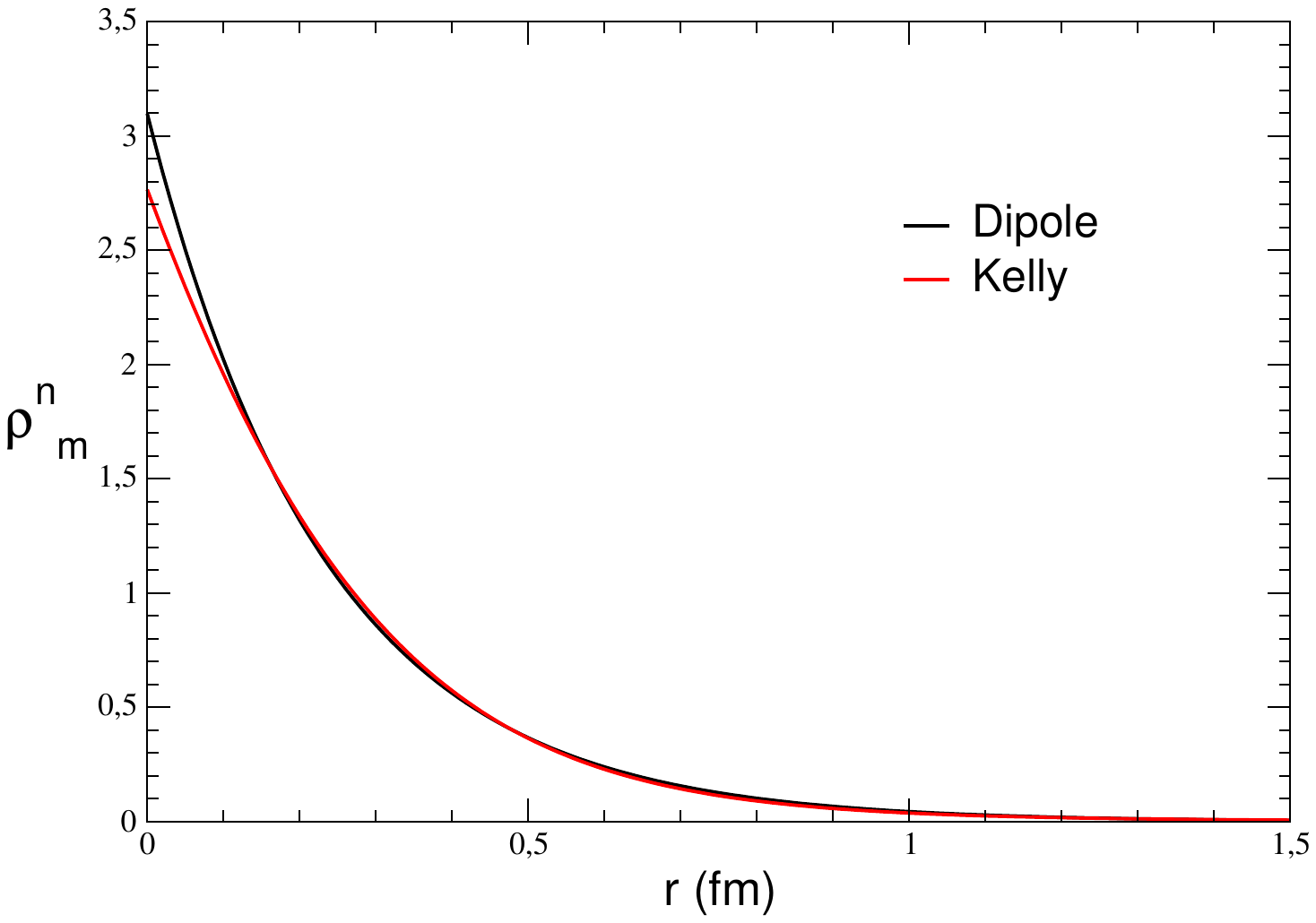}
\centering\includegraphics[width=8.cm]{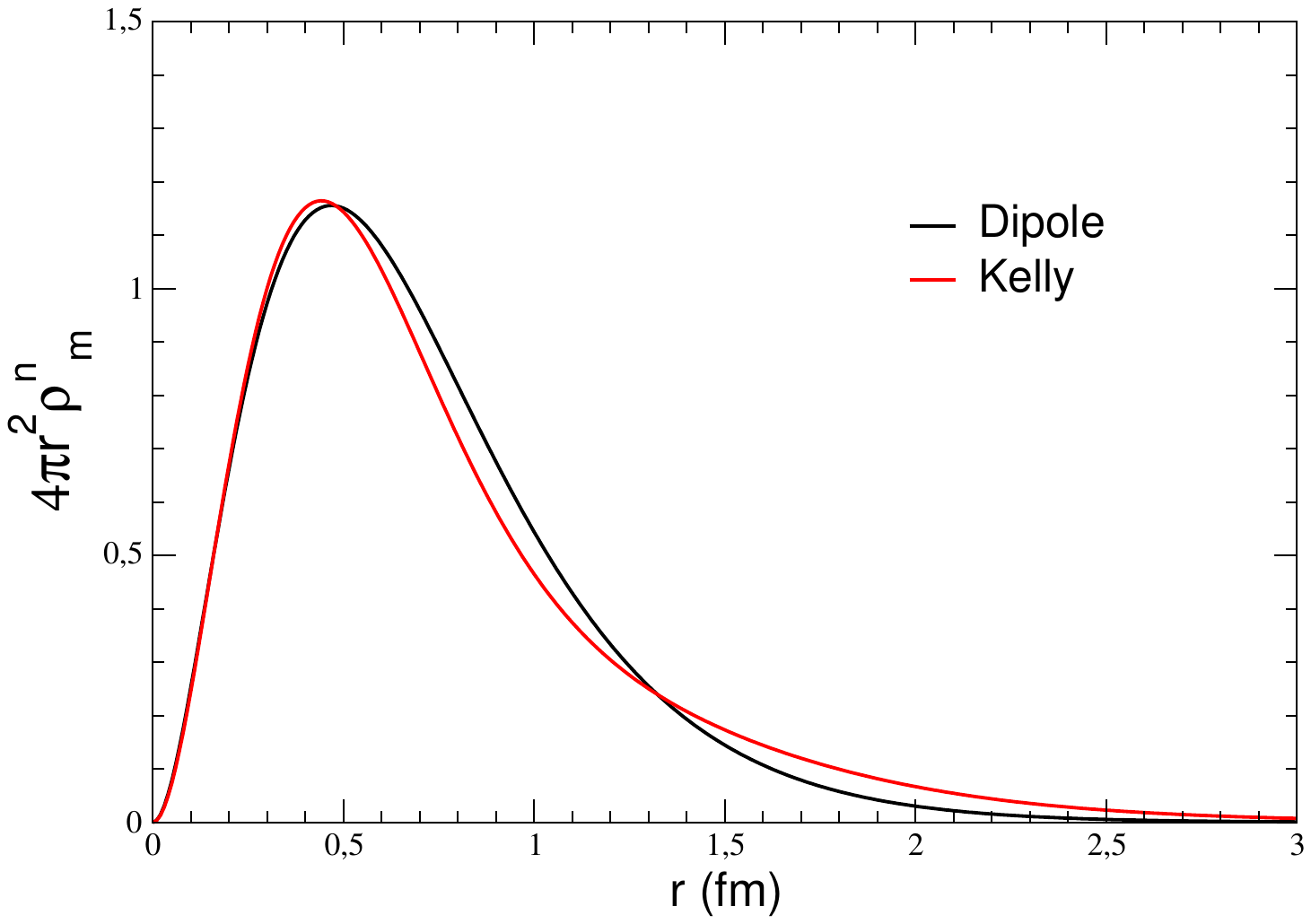}
\caption{Neutron magnetic densities (in $\mu_n$ units): $\rho^m_n(r)$ (upper panel)  and $4\pi r^2\rho^m_n(r)$ (lower panel),
  {obtained with different $n$ magnetic form factors: Dipole (\ref{GMn_Dipole}) and Kelly (\ref{GMn_Kelly})}.}\label{Fig_rhom_n}
\end{figure}

The $n$-magnetic densities $\rho^n_m(\vec{r})$ are obtained by Fourier transforming the $n$-magnetic form factor $G^n_M$ and must fulfil
\[ \int d\vec{r} \;  \rho_m^n(\vec r)=\mu_n\, , \]
where $\mu_n$ is the neutron magnetic moment in Bohr magneton units  $\mu_n$=-1.91304.  

By assuming a {    dipole} form  for the magnetic form factor~\cite{GKMSWB_1971,FN_ANP8_1975}
\begin{equation}\label{GMn_Dipole}
G_M^n(q^2) ={\mu_n\over \left( 1+ {q^2\over b_n^2} \right)^2}\, , 
\end{equation}
 the  $n$-magnetic density reads
\begin{equation}\label{rhom_n}
\rho_m^n(\vec{r})  = \mu_n\; {b_n^3\over 8\pi} e^{-x} \, .  
\end{equation}

We have also considered the more elaborate parametrisation of {    Kelly}, which reads:
\begin{equation}\label{GMn_Kelly}
  G^n_M(Q^2)= \mu_n  \frac{ 1+ a_1 \tau  }{ 1+ b_1\tau +  b_2 \;\tau^2 +  b_{3}\; \tau^{3}  }  \, , 
\end{equation}
with $\tau={q^2\over 4m_p^2}$, and involve  four dimensionless parameters:  $a_1=2.33 \pm 1.4$, $b_1= 14.72 \pm 1.7$, $b_2= 24.20 \pm 9.8$ and  $b_3= 84.1 \pm 41$.
The corresponding $n$-magnetic densities are depicted in Fig.~\ref{Fig_rhom_n} in $\mu_n$ units.
As one can see, the results for the magnetic density are more stable than for the charge density.

\section{The lepton-neutron interaction}\label{Sect_Vln}

We will consider on the same footing the three elementary leptons  ($e$,$\mu$ and $\tau$) that will be  generically denoted by $l=e^-,\mu^-,\tau^-$,
as well as their corresponding antiparticles $\bar{l}=e^+,\mu^+,\tau^+$.
The masses ($m_l$) are taken as $m_e$=0.510999\,MeV, $m_{\mu}$=105.658\,MeV and $m_{\tau}$=1776.86\,MeV,
and we will assume for all of them a Land\'e factor $g_l=2.00232$ \footnote{The  $g_l$ values for $e$ and $\mu$ are very close to each other and $g-2$ for the $\tau$ is poorly known}, such that their magnetic moments $M_l$  are given by 
\begin{eqnarray}
\label{M_l}
\vec{M}_l&=&g_l {q_l\hbar\over 2m_l} \vec{S} = \mu_l\vec\sigma\, ,\nonumber \\ 
\mu_l&=&-{g_l\over2} {e\hbar\over 2m_l} =-1.00116 \left({m_e\over m_l}\right) \;\mu_B\, , \nonumber\\
\quad \mu_B&=&  {e\hbar\over 2m_e} =5.78 83 82 \times10^{-5} \;{\rm eV \;T}^{-1}\, , 
\end{eqnarray}
where we  denoted $q_l=-e$, $q_{\bar{l}}=+e$, and $e$ is  the (positive) elementary charge.

For the neutron we  have taken $m_n$=939.565\,MeV, a Land\'e factor $g_n=-3.82608$  and a magnetic moment given by
\begin{eqnarray}\label{M_n}
&&\vec{M}_n=g_n {e\hbar\over 2m_p} \vec{S}= \mu_n\vec\sigma\, ,  ~     
\mu_n={g_n\over2} {e\hbar\over 2m_p}  =  -1.913 04 \; \mu_N\, , \nonumber \\
&&\mu_N= {e\hbar\over 2m_p} = 3.152451 \times10^{-8}  \; {\rm eV\; T}^{-1}\, .
\end{eqnarray}
For the remaining constants, we have taken the values  $\hbar c$=197.327\,MeV\,fm,  $1/\alpha$=137.036.

\bigskip
The lepton-neutron ($ln$) interaction is assumed to be purely electromagnetic, which means that we have neglected any weak contribution.
The interaction potential  we have considered has three components:  the Coulomb interaction $V^C_{ln}$,  
the dipole magnetic term $V^{MM}_{ln}$ resulting from the interaction between the
lepton and neutron magnetic moments and the spin-orbit term $V^{LS}_{ln}$.
\begin{equation}\label{Vln}
V_{ln}= V^C_{ln}+ V^{MM}_{ln}  +   V^{LS}_{ln}\, .
\end{equation}
The fist term ($V^C$)  is the purely Coulomb interaction between the pointlike lepton $l$ and the $n$ charge distribution.
The last two terms correspond to the Hyperfine Hamiltonian, as described e.g. in \cite{Jackson_CED_2001,Hagop_JMP351994},  integrated over the magnetization densities.
We neglect  $n$ polarization effects which, due to  virtual excitations to negative parity states,  could lead to a $1/r^4$ potential 
with a rich phenomenology of bound and resonant states, like in Refs. \cite{LC_FBS31_2002,CLDHK_EPL64_2003,CLK_JPB37_2004,DDB_PRA108_2023}. 

Each of the $V_{ln}$ terms depicted in \eqref{Vln}  are detailed in the coming sections.

\subsection{Coulomb interaction}

\begin{figure}[htbp]
\includegraphics[width=8.cm]{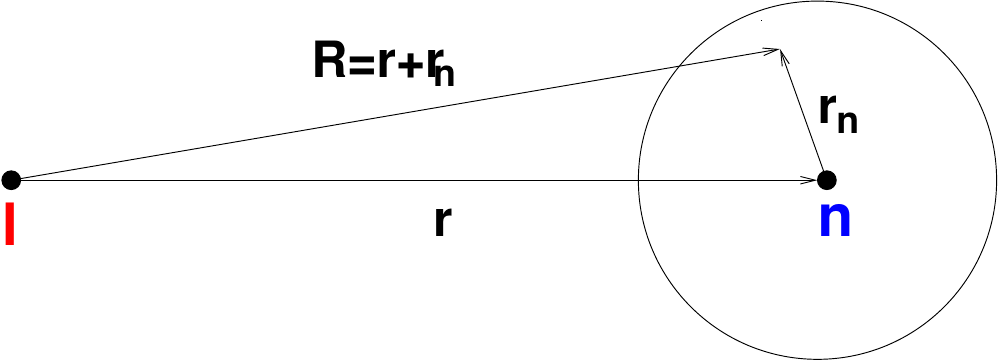}
\caption{Interaction between point-like lepton $l$  and $n$ charge distribution.}\label{Fig_ln}
\end{figure}

The $ln$ Coulomb interaction is entirely due to the $n$ internal structure.  It is obtained as a convolution of the Coulomb
potential between the point-like lepton with the $n$ charge density (see Fig.~\ref{Fig_ln}):
\begin{eqnarray}
V^C_{ln}(r)&=&- {1\over4\pi\epsilon_0} \int d\vec{r}_n  {e^2\rho_c^n(\vec{r}_n)\over \mid \vec{r}+ \vec{r}_n  \mid} \nonumber \\
&=& -\alpha(\hbar c)   \int d\vec{r}_n  { \rho^n_c(\vec{r}_n)\over \mid \vec{r}+ \vec{r}_n \mid}\, .
\end{eqnarray}

By inserting the Friar electric form factor~(\ref{GEn_F}) in the previous expression and making use of
\[ {1\over R} = {1\over 2\pi^2} \int d\vec{q}  \; {e^{i \vec{q}\cdot\vec{R} } \over q^2}\, , \]
the lepton-neutron Coulomb potential reads:
\begin{equation}\label{VCln}
\hspace{-.2cm} V^C_{ln}(r) =  -\;  \alpha(\hbar c)  \: b_n  \;   {(\beta_nb_n^2)\over 8} \;  \left(  1 + x \right)  \;  e^{- x}\, ,  ~ x =b_nr \, .
\end{equation}
In the point-like  limit, $\beta_n\to 0$, and therefore the potential vanishes.

\begin{figure}[htbp]
\includegraphics[width=8.cm]{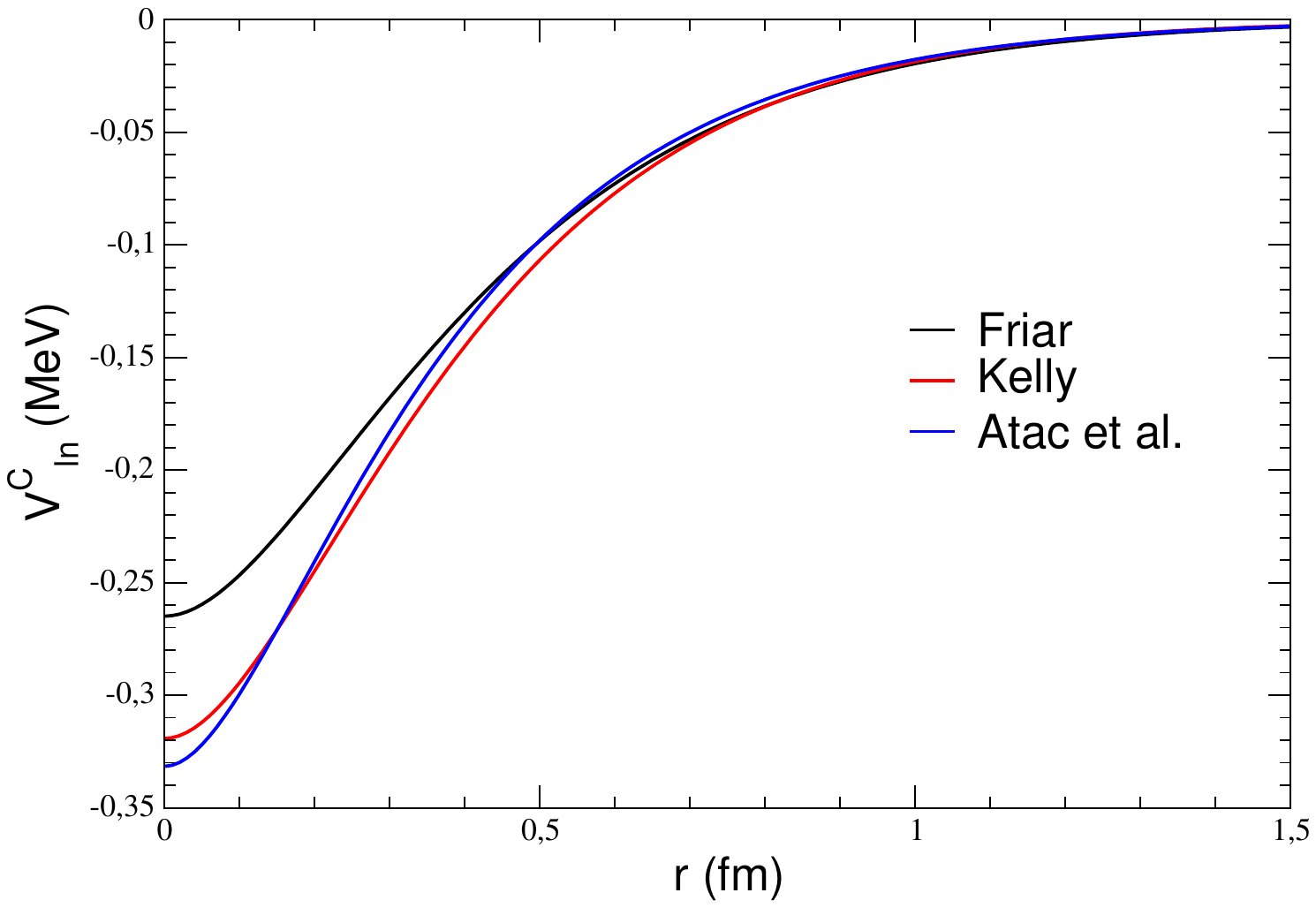}
\includegraphics[width=8.cm]{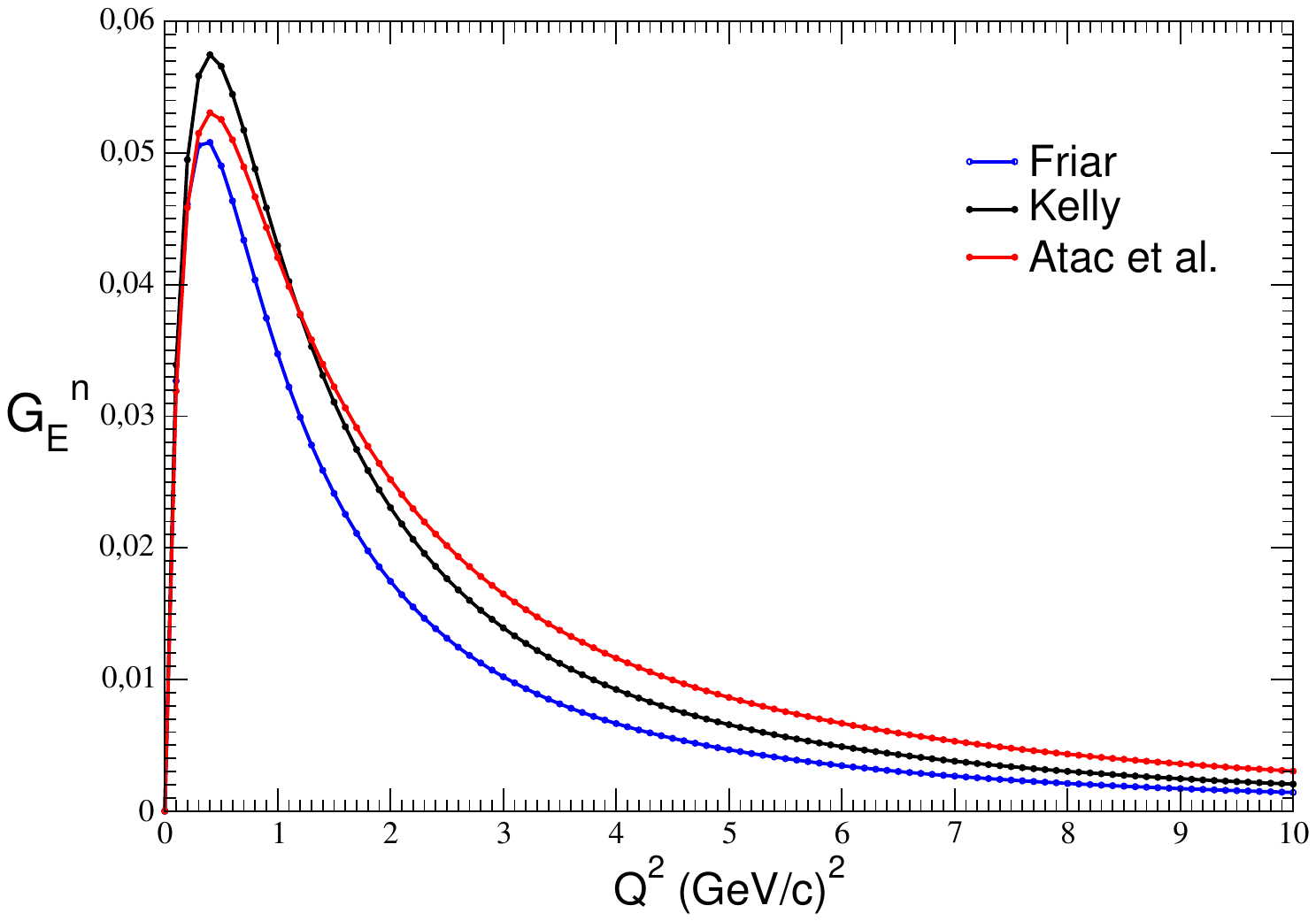}
\caption{Upper panel: Coulomb potential between a lepton $l$ and a neutron $n$ (in MeV and fm units) with (lower panel)  the corresponding $n$ electric form factors,
  { Friar (\ref{GEn_F}), Kelly  \ref{GEn_Kelly})  and Atac et al. \cite{Atac_Nature_Com_2021}, used in their computation. The potential is the same for all leptons.}}\label{Fig_VCln}
\end{figure}

This potential, {    which is the same for the three leptons}, is displayed in the upper panel of Fig.~\ref{Fig_VCln}  (solid black line) in MeV and fm units.
It is monotonously attractive with a depth at 
the origin of $V^C_{ln}(0)\equiv C^C_{ln}\approx-0.266$ MeV.
We have also included for comparison the results obtained with other parametrisations of the $n$ charge density represented in the lower panel: the original Kelly parametrisation from~\cite{Kelly_PRC70_2004}	(in red) and the recent  readjustment of the Kelly parameters from~\cite{Atac_Nature_Com_2021} (in blue).
Their analytic expressions are quite lengthy and are omitted here.
The noticeable differences observed in the $n$ charge densities are also manifested in the $ln$ Coulomb potentials at $r\approx0$.

Notice that for the corresponding antiparticles (in our convention with positive charge), the sign of  potential~(\ref{VCln}) must be changed, giving rise to different
$ln$  and $\bar{l}n$ low energy parameters. This difference -- at first glance surprising since dealing with scattering on a neutral particle  -- 
is uniquely due to the neutron's internal structure.

\subsection{Magnetic dipole interaction}

The interaction between two point-like magnetic  moments is given by~\cite{Jackson_CED_2001,Hagop_JMP351994}:
\begin{multline}
\label{VMM_EW2}
V_{MM}(\vec{r}) = -{\mu_0\over4\pi} \left[  {8\pi\over3}\;  \vec{M}_l\cdot \vec{M}_n  \;\delta(\vec{r}) \right. \\ \left. \;+\;  \frac{ 3 (\vec{M}_l\cdot \hat{r}_l)(\vec{M}_n\cdot \hat{n}) -\vec{M}_l\cdot \vec{M}_n }{ r^3} \right]\, ,
\end{multline}
which, in terms of (\ref{M_l}) and (\ref{M_n}),  can be written as
 \begin{equation}\label{VMM_EW4}
\hspace{-.3cm}  V_{MM}^{ln}(\vec{r})=-{\mu_0\mu_l\mu_n\over4\pi} \; \left[{8\pi\over3}\; \vec{\sigma}_l\cdot\vec{\sigma}_n \;\delta(\vec{r}) \;+\;  \frac{ \hat{S}_{12}(\hat{r})  }{ r^3} \right] \, ,
\end{equation}
where
\(  \hat{S}_{12} (\hat{r}) \equiv 3 (\vec{\sigma}_1\cdot \hat{r})(\vec{\sigma}_2\cdot \hat{r}) -\vec{\sigma}_1\cdot \vec{\sigma}_2 \)
is the tensor-operator, whose matrix elements are given by
 \begin{multline}\label{S12}
   \langle  SLJ\mid S_{12} \mid S'L' J'\rangle= \delta_{SS'}\delta_{S1}\delta_{JJ'} \; \\ \times
    \bordermatrix{& L=J-1                               & L=J&  L=J+1 \cr
                  {\rm L=J-1} &  {-2L\over2L+3}                &   0   & {6\sqrt{J(J+1)}\over2J+1} \cr
                         L=J      &       0                                &   2   &        0                   \cr
                         L=J+1  & {6\sqrt{J(J+1)}\over2J+1} &   0   &     {-2(L+1)\over2L-1}       \cr}   
\end{multline}
  {and
\begin{equation}\label{sigma.sigma}
\hspace{-0.3cm} \langle  SLJ\mid \vec\sigma_1\cdot\vec\sigma_2\mid S'L' J'\rangle=(-3\delta_{S0}+\delta_{S1})   \delta_{SS'} \delta_{LL}\delta_{JJ'}.
\end{equation}
}

In order to take into account  the $n$  magnetization density, the expression for the $n$ magnetic moment  (\ref{M_n}) becomes:
\[   \vec{M}_n= \mu_n \int \rho_m^n(\vec{r}) \vec{\sigma} \; d\vec{r}  \]
and Eq.~(\ref{VMM_EW4}) is generalized into
\begin{multline}\label{VMM_rho}
  V_{MM}^{ln}(\vec{r})=-{\mu_0\mu_l\mu_n\over4\pi}
 \left[  {8\pi\over3}\; \vec{\sigma}_l\cdot\vec{\sigma}_n  \int d\vec{r}_n \rho_m^n(\vec{r}_n )  \delta(\vec{R})\right.
 \\ \left. +  
  \int d\vec{r}_n  
  \frac{ 3 (\vec{\sigma}_l\cdot \hat{R})(\vec{\sigma}_n\cdot \hat{R}) -\vec{\sigma}_l\cdot \vec{\sigma}_n  }{ R^3} \;\rho_m^n(\vec{r}_n) \right]   \,, 
\end{multline}
where $\vec{R}=\vec{r}+\vec{r}_n$.

By inserting the {    dipole form factor} (\ref{GMn_Dipole}),  the integration can be performed analytically,  as for the Coulomb case,
and the $ln$ magnetic interaction reads:
\begin{multline}\label{Vln_MM}
 V_{MM}^{ln}(x)= - {\mu_0\mu_l\mu_n\over 4\pi} \; b_n^3 \; \left[     {1\over3} e^{-x}  (\vec\sigma\cdot \vec\sigma) \right. \\ \left. + \frac{ 1- \left( 1 + x + {x^2\over2}  + {x^3\over6}  \right) e^{-x}}{ x^3}\; \hat{S}_{12} \right]\, ,
\end{multline}
 where $ x=b_n r$.
 
By  writing explicitly the scalar spin-spin ($V_S$) and tensor ($V_T$) components we can write~(\ref{Vln_MM}) in the form:
\begin{equation}\label{Vln_Op}
  V_{MM}^{ln}(x)=      V_{S}(x)  (\vec\sigma\cdot \vec\sigma)  + V_{T}(x)   \hat{S}_{12} \, , 
\end{equation}  
where
\begin{eqnarray}
V_S(x)&=& - {\; 1\over3} \;  C_{MM}^{ln}\;     e^{-x}\, , \\\
V_T(x) &=& - C_{MM}^{ln} \;   \frac{ 1- \left( 1 + x + {x^2\over2} + {x^3\over6}  \right) e^{-x}}{ x^3}\, ,
\end{eqnarray}
and the (positive) numerical pre-factor
\begin{equation}\label{Cln_MM}
C_{MM}^{ln}=\mu_0 {\mu_n\mu_l\over 4\pi} \; b_n^3 
=\;  -\;{g_ng_l\over4}  \;  {\alpha  \;(b \hbar c)^3\over 4(m_nc^2)(m_lc^2)}\, .
\end{equation}
For the  electron case  ($l=e^-$) it takes the value $C_{MM}^{en}$=\,4359.4109\,MeV.

\newcommand\egal{\mathop{\approx}}

Notice that the $ln$ magnetic potential~(\ref{Vln_Op}) for  different leptons differs from each other only by the value of this pre-factor, which merely scales the respective 
($V_S$) and ($V_T$) components. In view of further  discussions, it is interesting  to take as a reference the $en$ case and write 
\begin{equation}\label{Vln_Scale}
   V_{MM}^{ln}(x) = \left( {m_e\over m_l}\right)  \; V_{MM}^{en}(x)  \, .
\end{equation}   

\begin{figure}[htbp]
\includegraphics[width=8.5cm]{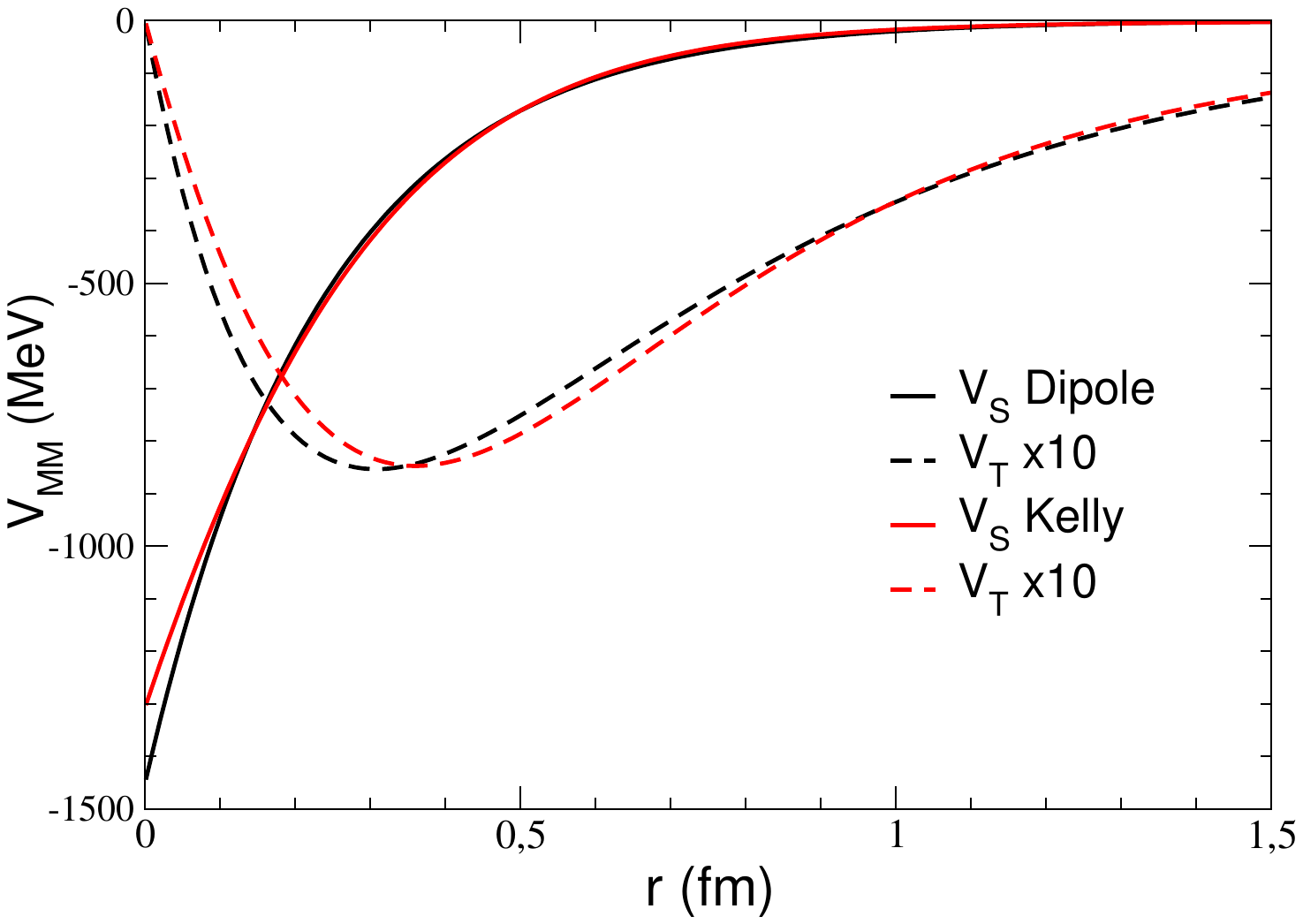}\vspace{-0.3cm}
\includegraphics[width=8.5cm]{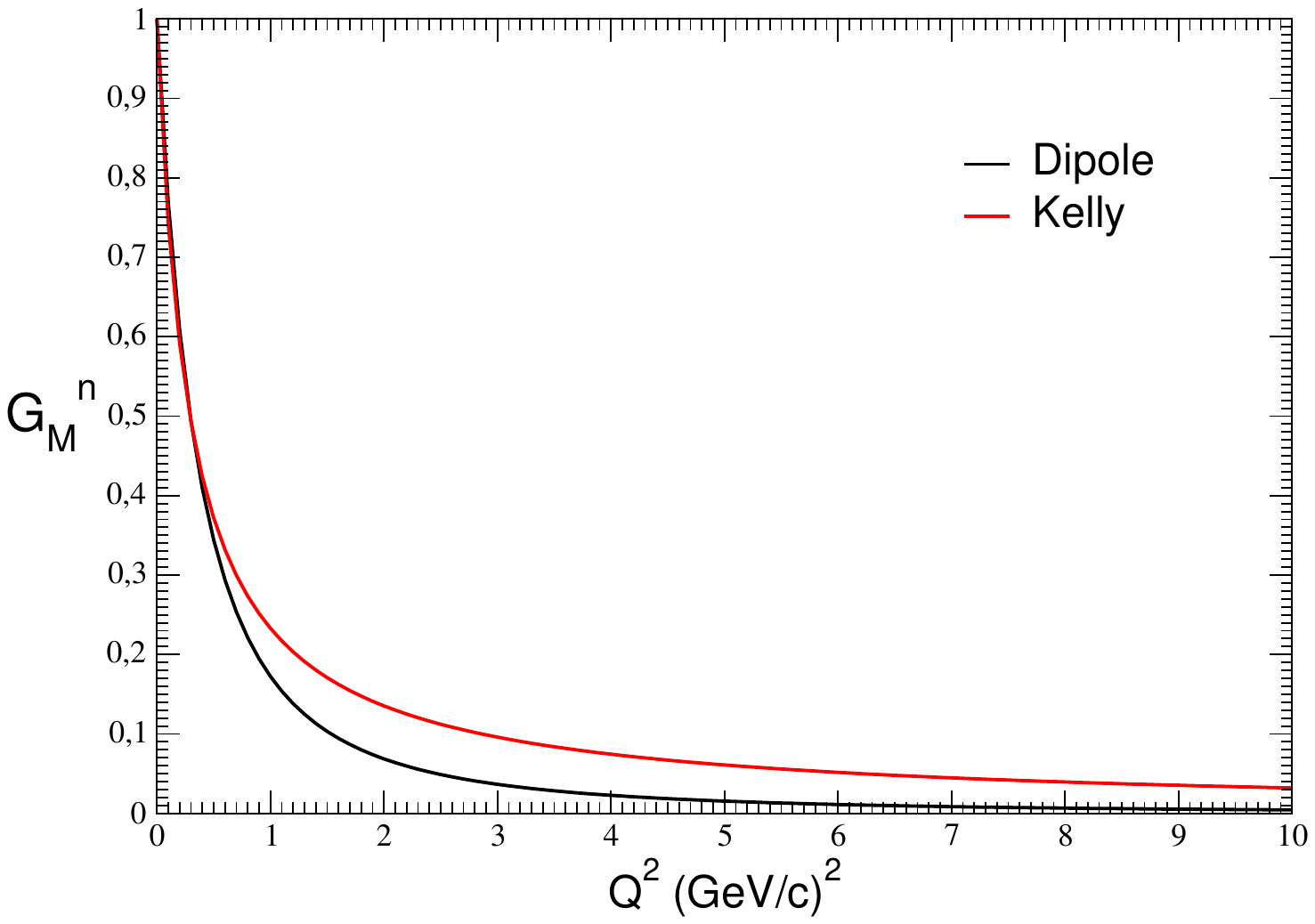}
\caption{Upper panel : Spin-spin ($V_S$) and tensor ($V_T$) components of the  magnetic interaction   {(\ref{Vln_Op})} between  $e^-$  and $n$, corresponding to the Dipole (\ref{GMn_Dipole}) and Kelly  (\ref{GMn_Kelly})  magnetic form factors ($G_M^n$),    {which are} represented in the lower panel.}\label{Fig_VMMen}
\end{figure}

We have displayed in  Fig.~\ref{Fig_VMMen} the spin-spin ($V_S$) and tensor ($V_T$) components of the reference  magnetic potential $V_{MM}^{en}$.
As one can see, $V_S$  largely dominates at small distance, where it takes values as large as 1.5\,GeV;  at $r$=0.5\,fm  one still has $V_S\approx 200$\, MeV.
Due to the finite size structure of $n$, both components are finite in all the domain $[0,+\infty]$ and $V_T$ has the asymptotic  behaviours
\begin{equation}\label{VT_Ass}
 V_T(x) \egal_{x\to0} -\;C_{MM}^{ln}  \; {x\over24}\,,  \qquad  V_T(x) \egal_{x\to\infty} -\;{C_{MM}^{ln}  \over x^3} \, ,  
 \end{equation}
with a maximum at $r\approx0.4$\,fm.
We have  also included in Fig.~\ref{Fig_VMMen}  the $V_{MM}^{ln}$ potential  provided the {    Kelly magnetic form factor} (\ref{GMn_Kelly}). The result is still analytic
but the expression is lengthy enough to be omitted in the text.
As was the case for the Coulomb interaction $V_C$, $V_S$ displays some sizeable differences at $r=0$ among the models.

Notice that for the $en$ case,  the Coulomb potential~(\ref{VCln}), displayed in Fig.~\ref{Fig_VCln}, is totally negligible with respect to the magnetic one~(\ref{Vln_Scale}).
However, while the former is independent of the lepton flavour,  the latter one scales with the inverse of lepton mass and the situation
is reversed in  the case of $\tau$.

In view of the sizeable values of the spin-spin component $V_S$, the question of a possible $en$ bound state seems, a priori, pertinent and will be examined in the next section.
However, the value of ${\hbar^2/(2\mu_{ne})}\approx38120$\,MeV\,fm$^2$,  driving the repulsive kinetic energy term, lets very little hope for the $en$ case.
At $r\approx 0.8$\,fm the slow-decreasing tensor component starts being dominant and its $1/r^3$ tail imposes non trivial asymptotic 
conditions for the scattering solutions in the spin-triplet (S=1) $L>0$ states, for which the standard LEPs are not defined.

\subsection{Spin-orbit interaction}

Our starting point is the  spin-orbit term of the Hyperfine interaction 	for a point-like lepton \cite{Jackson_CED_2001,Hagop_JMP351994}:
\begin{equation}\label{Hhyp_eN}
H^{ln}_{LS}  = \;  {\mu_0\over4\pi}   {e\over m_e}  {1\over r^3} \mathbf{L}\cdot \vec{M}_n     = \;-\;  {\mu_0\over4\pi}   {e\mu_n\over m_e} \; \vec{\sigma}_n  \cdot   {(\mathbf{p}\wedge \mathbf{r}) \over r^3} \,.
\end{equation}
If one takes into account the $n$ magnetization density, this expression generalizes into
\begin{equation}\label{Hhyp_erho}
H^{ln}_{LS}  = \; - {\mu_0\over4\pi}   {e\mu_n\over \mu_{ln}}  \;    \vec{\sigma}_n  \cdot   \mathbf{p}\wedge  \int d{\mathbf r}_n \;    { {\mathbf R}  \rho^n_m(  {\mathbf r}_n )  \over R^3}\,,
\end{equation}
where we have used the notation of Fig.~\ref{Fig_ln}.  In principle, an additional  term should be added to (\ref{Hhyp_erho})
to account for the coupling between the $e$ magnetic moment and the magnetic field created by the orbiting $n$.
The non-zero $n$ charge density will indeed create a current and the corresponding magnetic field. This term is supposed to be negligible
and has been omitted.

By using the same techniques developed for the charge and magnetic terms, one obtains for the spin-orbit interaction the general form
\begin{equation}\label{Vln_LS}
V^{ln}_{LS}= V_{LS} (x) \; (\vec L\cdot \vec{s}_n )\,.   
\end{equation}
 
\begin{figure}[htbp]
\centering\includegraphics[width=8.5cm]{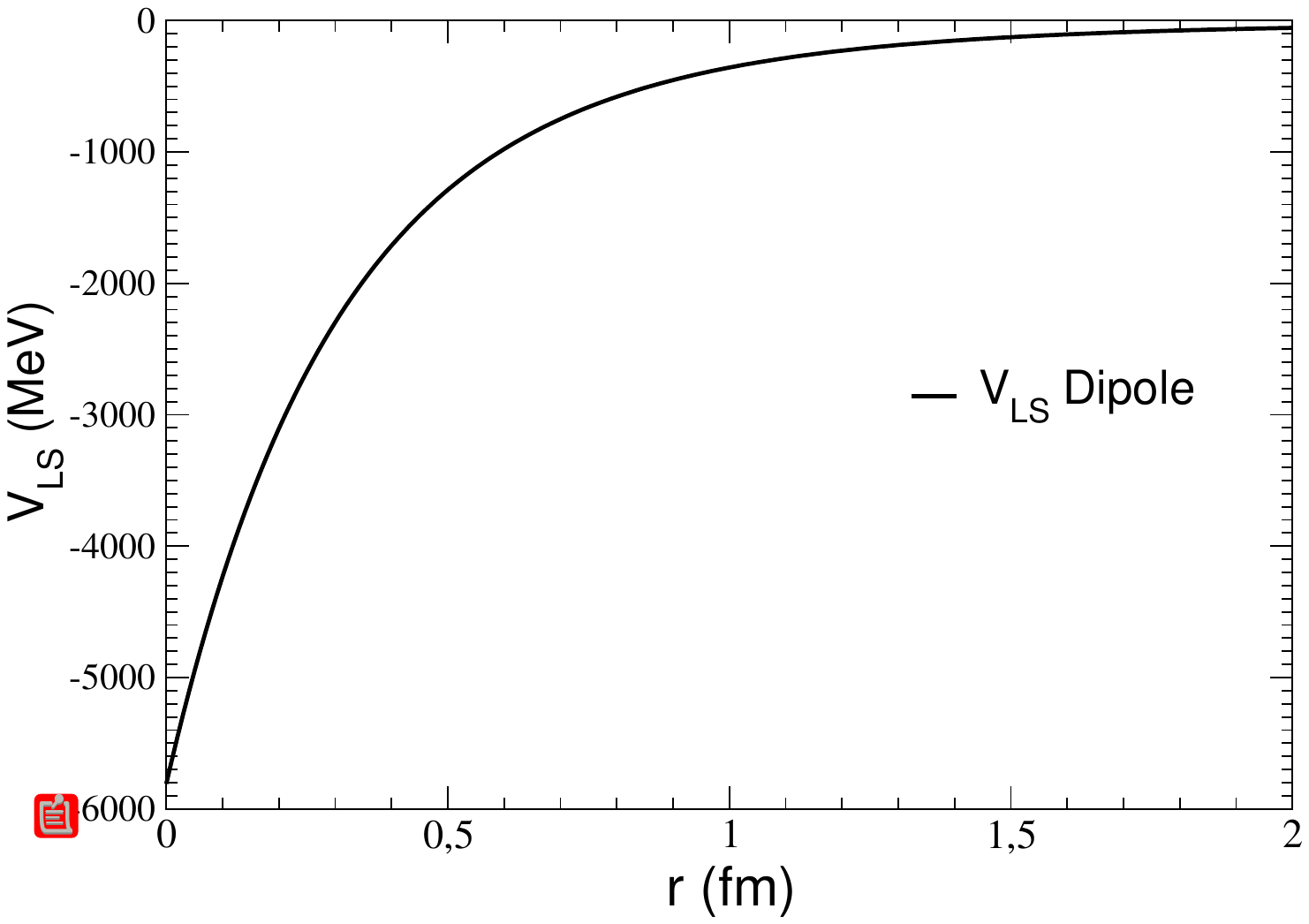}
\caption{Spin-orbit potential  ($V_{LS}$)  for the $en$ scattering   {obtained with the  Dipole $n$ magnetic form factor (\ref{GMn_Dipole})}.}\label{Fig_VLS_en}
\end{figure}

When inserting the dipole form factor one has
\begin{equation}\label{VLS_F}
V_{LS}(x) = C^{ln}_{LS} \;   \frac{ 1 -  (1+x + {x^2\over2}) e^{-x}     }{x^3}  
\end{equation}
with
\begin{small}
\begin{equation}\label{VLS_Ass}
\hspace{-.3cm} V_{LS} (x) \egal_{x\to0} C^{ln}_{LS}  \left[ {1\over 6} - {x\over 8} + O(x^2)  \right] \,,  ~  V_{LS}( x) \egal_{x\to\infty} {C^{ln}_{LS}  \over x^3} \, ,  
 \end{equation}
 \end{small}
and $ C^{ln}_{LS}$ a (negative, since $g_n=-3.8261$) numerical pre-factor 
\begin{equation}\label{Cln_LS}
 C^{ln}_{LS} = g_n\;  {\alpha (b\hbar c)^3 \over    (\mu_{ln} c^2)( m_p c^2)} \,\,.   
 \end{equation}
For the $en$ case one has $C^{en}_{LS}=  -34853,82$\,MeV. 
The corresponding potential is displayed in Fig.~\ref{Fig_VLS_en}.
As for the spin-spin term, there is a deep attraction at the origin  but it is compensated by the centrifugal barrier in such a way that the effective potential is repulsive everywhere.
Remarkably, the reduced spin-orbit potential (i.e. $v_{LS}=2\mu_{ln}V_{LS}$) is the same for the three leptons.

\newcommand{\sixj}[6]{\left\{\begin{array}{ccc} #1&#2&#3\\#4&#5&#6 \end{array}\right\}}

Notice that the total orbital angular momentum $L$ of the $ln$ pair is not coupled to its total spin $S=s_n+s_e$ but only to the neutron spin $s_n$.
In this sense, the interaction does not correspond to the standard spin-orbit interaction, although we will keep the same notation to denote it.
The main difference is that interaction (\ref{Vln_LS}) does not conserve the total spin $S$ 
 in a similar way that the tensor term does not conserve $L$.  
 The matrix elements of the spin-orbit operator (\ref{Vln_LS}) in the standard partial wave basis  $\mid SLJ\rangle$ are:
 \begin{itemize}
 \item  Null  for S-waves
  \begin{equation}
 \langle  ^1S_0           \mid   \vec{L} \cdot \vec{s}_n   \mid ^1S_0\rangle  = \langle  ^3S_1            \mid   \vec{L} \cdot \vec{s}_n  \mid ^3S_1\rangle   =  0  
 \end{equation}
 \item For $L>0$ triplet unnatural parity states
 \begin{equation} \label{Lsn_SLJ_UPS}
\langle   ^3L_{L\pm1} \mid   \vec{L} \cdot \vec{s}_n  \mid ^3L_{L\pm1}\rangle =\lambda_{\pm}(L) 
\end{equation}
with $\lambda_{\pm}$ given in \eqref{lambda_pm}.
\item They couple the $L>0$  singlet and triplet natural parity states
\begin{multline}\label{Lsn_SLJ_NPS}
 \langle  ^{2S+1}L_{J=L}\mid   \vec{L} \cdot \vec{s}_n \mid ^{2S'+1}L_{J=L} \rangle
 =\\ =\begin{bordermatrix} { &    S=0          &   S=1            \cr
                               S=0 &         0          & \sqrt{L(L+1)} \cr 
                               S=1 & \sqrt{L(L+1)}& -1                 \cr}\end{bordermatrix}
\,.\end{multline}
\end{itemize}
Their computation requires some care and it is detailed in the Appendix \ref{Ap_A}

\section{Results}\label{Sect_Results}

We  present in this Section  the scattering results obtained with the above detailed $V^{ln}$ potentials,
for  some selected $ln$ states. To this aim, we write the total potential in the operator form
\begin{multline}
\label{VTOT}
 V^{ln}(r)= V^{ln}_C(r) + V^{ln}_{S}(r)  \;(\vec\sigma_l\cdot \vec\sigma_n)\\  + V^{ln}_{T}(r)  \; \hat{S}_{12}  + V^{ln}_{LS}(r) \;(\vec{L}\cdot\vec{s}_n ) \, .
\end{multline}
It depends on four scalar functions $V^{ln}_{i=C,S,T,LS}$ which change their sign for the antilepton scattering: $V^{\bar ln}_i=-V^{l n}_i$.

Due to the tensor and spin-orbit terms, the physical states are in general labeled  only by $J^{\pi}=0^{\pm},1^{\pm},2^{\pm}...$ quantum numbers with $\pi=(-)^L$. 
Calculation are performed in the $\mid SLJ\rangle$ basis  and we will  use the spectroscopic notation  $^{2S+1}L_J$ for the tensor and spin-orbit uncoupled states,
the standard  notation $^{2S+1}$L$_J$-$^{2S+1}$(L+2)$_J$ for the tensor coupled ones, and the $^1$L$_L$-$^3$L$_L$ for the spin-orbit coupled states.

\begin{figure}[t]\vspace{-0.5cm}
\centering\includegraphics[width=8.5cm]{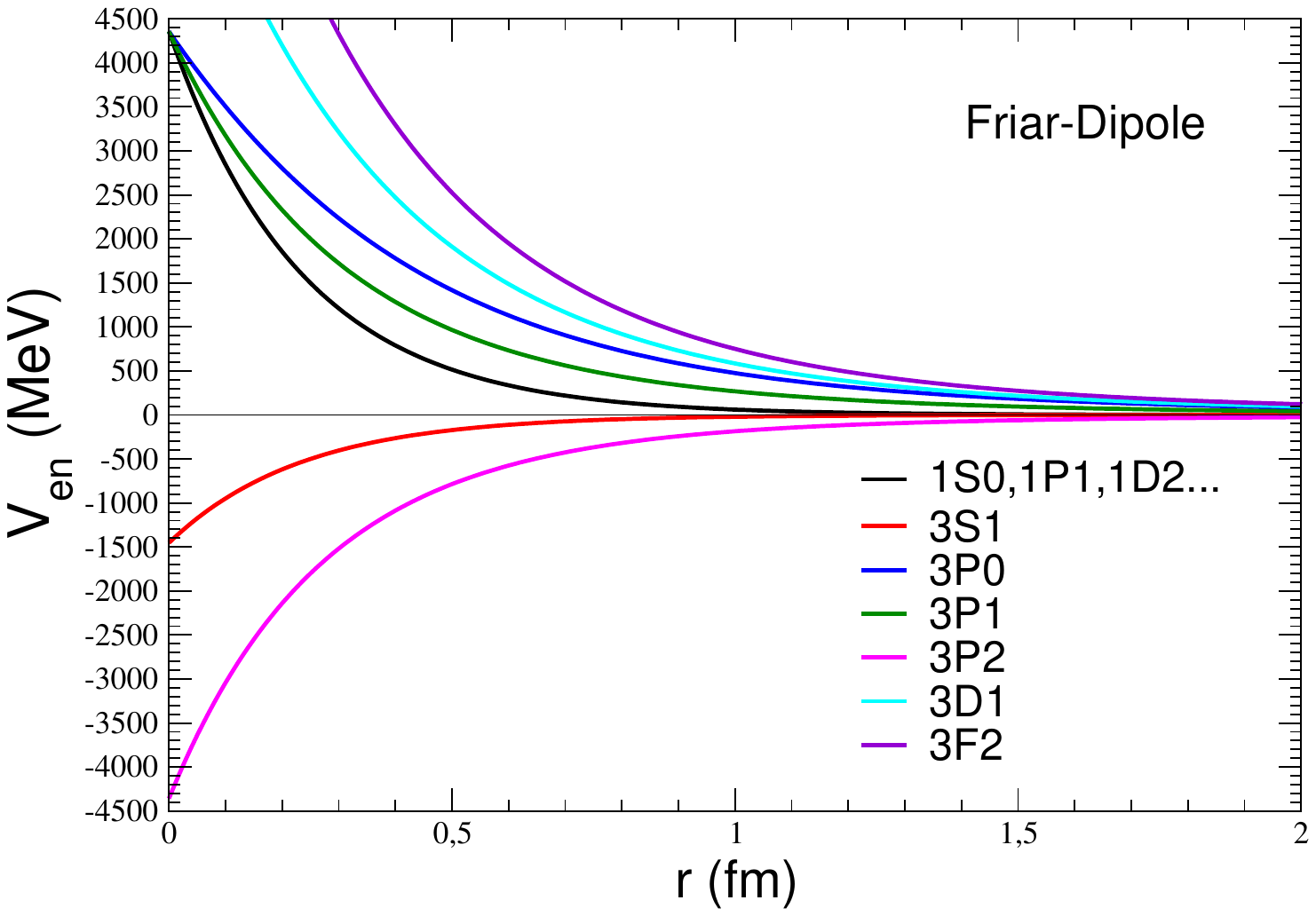}\vspace{-0.5cm}
\centering\includegraphics[width=8.5cm]{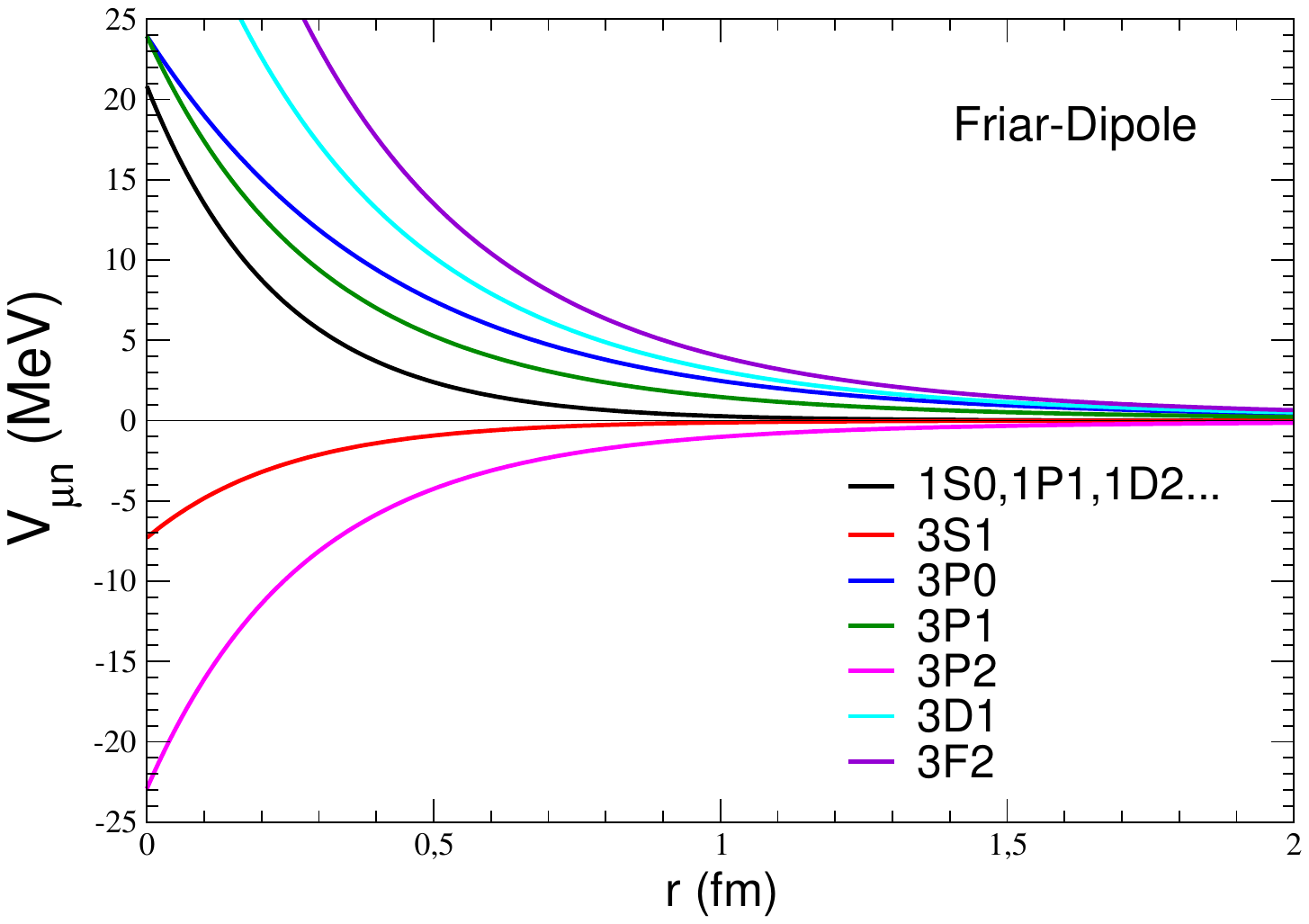}\vspace{-0.5cm}
\centering\includegraphics[width=8.5cm]{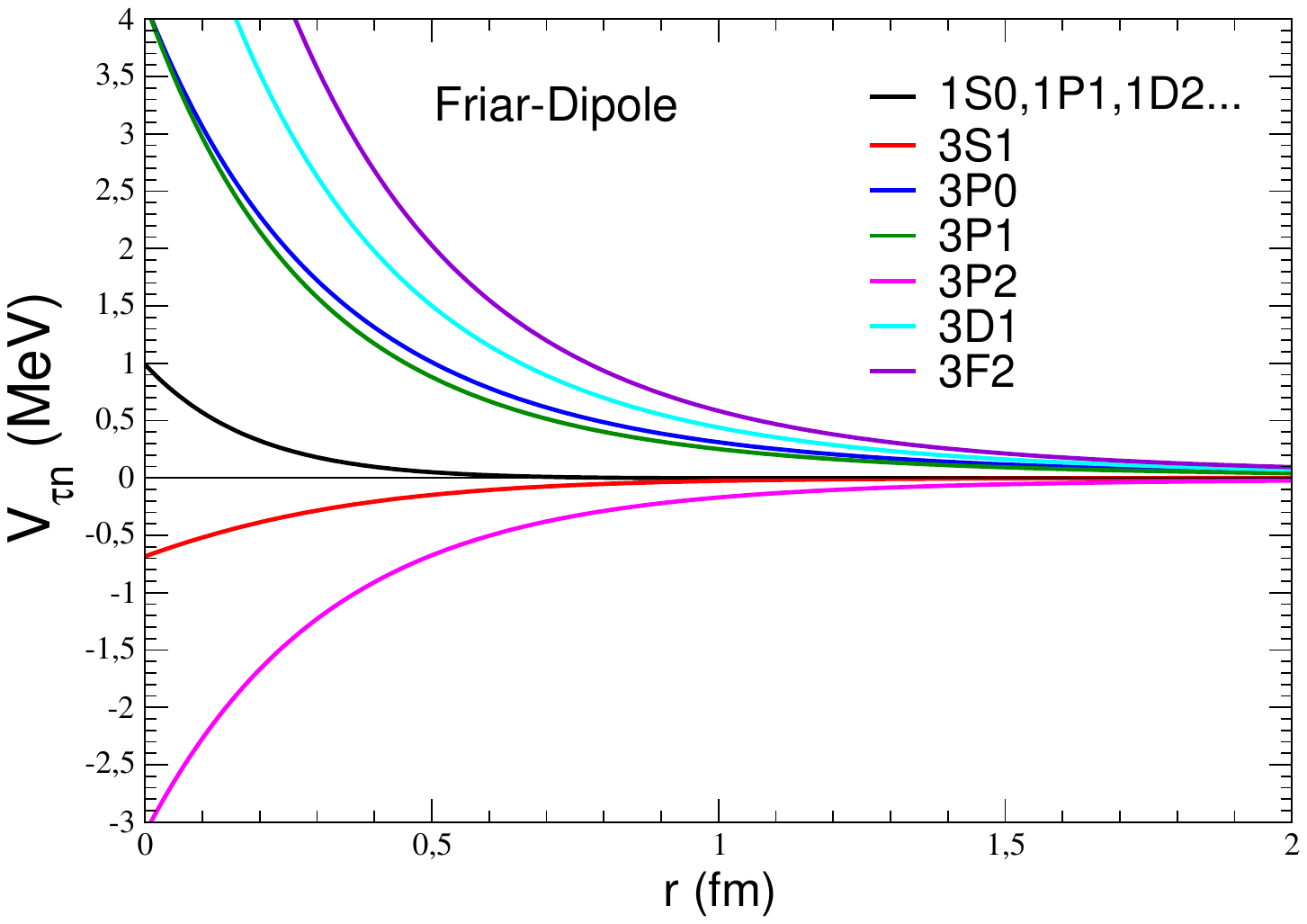}
\caption{$ln$ potentials  in different partial waves, obtained with the Friar  (electric)  and Dipole (magnetic) $n$ form factors.   {From top to bottom: $en$, $\mu n$ and $\tau n$}.}\label{Fig_Vln_FD}
\end{figure}

The matrix elements of the spin-spin, tensor and spin-orbit operators in this basis  are given in Table~\ref{Ops} for the lowest partial waves
and the corresponding $V^{ln}$ potentials  are displayed in Fig.~\ref{Fig_Vln_FD} for the three considered leptons (in MeV and fm units). 
Notice the different energy scales among them, varying from few MeV (for $\tau n$)  to few GeV (for $en$), which are essentially due to the involved magnetic moments.
The $V^{ln}$  potential is the same for all the singlet states ($^1$S$_0$,$^1$P$_1$,$^1$D$_2$,...), since the tensor and the diagonal term of  $(\vec{L}\cdot\vec{s}_n )$ vanishes.
All potentials are strongly repulsive, except the $^3$L$_{J=L+1}$ states ($^3$S$_1$ and  $^3$P$_2$ in the selected ensemble) 
which are attractive, in absence of the centrifugal term. 
Let us remind  that the situation is however reversed for the antilepton-neutron cases.
  {Notice that  for $en$ case, there is a merging of $^1$S$_0$, $^3$P$_0$ and $^3$P$_1$ potentials at $r=0$, and	
that the $^1$S$_0$  result is getting away when going to $\mu n$  and $\tau n$. 
The reason for that  lies in the particular expressions of potentials and the angular matrix elements presented in Table \ref{Ops}.
The expressions for these potentials are given below:  
\begin{eqnarray}
V^{ln}_{^1S_0} &=& V^{ln}_C - 3 V^{ln}_{S},    \label{V1S0}  \\
V^{ln}_{^3P_0} &=& V^{ln}_C + V^{ln}_{S} -  4\; V^{ln}_T - V^{ln}_{LS},\qquad    \label{V3P0}  \\ 
V^{ln}_{^3P_1} &=& V^{ln}_C + V^{ln}_{S} +2\; V^{ln}_T - V^{ln}_{LS}.\qquad    \label{V3P1} 
\end{eqnarray}
At $r=0$, $V_T$ vanishes and $V^{ln}_{^3P_0}(0)=V^{ln}_{^3P_1}(0)$ for all leptons.
For the $en$ case, the equality between these three potentials at the origin is due to the approximate relation $8C^{en}_{MM}\approx C^{en}_{LS}$  (at the level of 0.1\%)
that follows  from Eqs. \eqref{Cln_MM}
and \eqref{Cln_LS} with $\mu_{en}\approx m_e$ and $g_e\approx 2$. This approximate relation is broken when the lepton mass increases from $e$ to $\mu$, and $\tau$
as clearly seen in the figure.}

\begin{small}
\begin{table}[t!]
\begin{center}\begin{tabular}{|   l  |   l   |    l   |   l l   |  l l   | }\hline
               &                                      & $\sigma_l\cdot\sigma_n$ & \;\;\;\; \;\;$S_{12}$ &          & \;\;\;$\vec{L}\cdot\vec{s}_n$ & \\\hline  
J$^{\pi}$ &$^{2S+1}L_J$                &   S     \;\;\;\;S'                    & L                 &  L'                &       S  &  S' \\\hline
$0^+$     &$^1$S$_0$                    &                                  -3    &   0                &                    &  0\;\;\; &  \\\hline
$1^+$     &$^3$S$_1$-$^3$D$_1$ &  +1                                  &    0               & 2$\sqrt{2}$  &   0\;\;\; & 0\\
&             &                                      &2$\sqrt{2}$                      & -2                 & 0 \;\;             & -3/2   \\\hline
$0^-$      &$^3$P$_0$                    &+1                                    & -4                 &                    &   -1 \;\; &        \\\hline
$1^-$      &$^1$P$_1-^3$P$_1$     & -3 \;\;\;\; 0                        &   0                &                    &  0 \;\;   & $\sqrt 2$\cr    
               &                                      &  0 \;\;\;\; 1                       &                    &     2               &  $\sqrt 2$ \;\;  &  -1 \\\hline
$2^-$     &$^3$P$_2$-$^3$F$_2$ & +1                                   &  -2/5              &6$\sqrt{6}$/5&  +1/2  \;\; & 0  \cr
              &                                      &                                        & 6$\sqrt{6}$/5 &   -8/5           &   0  \;\; &      -2 \\\hline      
\end{tabular}\end{center}
\caption{  {Angular matrix elements of the spin-spin (\ref{sigma.sigma}), tensor (\ref{S12}) and spin-orbit (\ref{Lsn_SLJ_NPS})} operators for the lowest partial waves.}\label{Ops}
 \end{table}
 \end{small}

These potentials are inserted in the set of coupled reduced radial Schrodinger equations
\[ \partial_r^2 \varphi_{\alpha}(r) + \left[   k^2 - \frac{L_{\alpha}(L_{\alpha}+1)}{r^2} \right] \varphi_{\alpha} (r)
= \sum_{\beta} v_{\alpha\beta} \varphi_{\beta}\,,  \]
where $ v_{\alpha\beta} = {2\mu_{ln} \over\hbar^2} \; V_{\alpha\beta} (r) $ is the reduced potential and $k^2={2\mu_{ln}\over\hbar^2}E$ is the center of mas momentum. 
Remarkably, the huge variations observed in Fig.~\ref{Fig_Vln_FD}  between the different leptons are largely compensated
by their reduced mass in $v_{\alpha\beta}$ and,  the resulting $ln$ scattering observables 
turn to be quite similar among them, especially in the zero energy limit.
This will be presented in the following subsections.

It is worth noticing that, except from the $^1$S$_0$ and $^3$S$_1$ states, all the partial waves potentials behave asymptotically as $1/r^3$,
due to both the  tensor and the spin-orbit terms. In this case, the standard scattering theory does not apply \cite{SOMR_PRL5_1960,SOMR_JMP2_1961,GG_NC38_1965_I,GG_NC40_1965_II,Shakeshaft_JPB5_1972,Gao_PRA59_1999,Muller_PRL110_2013}, in particular
the low-energy parameters are not defined and the low-energy limit of the cross sections is strongly modified.
For the singlet states both terms are absent but, as we have discussed in the previous section, they are
coupled by a long-range term to the triplet state, which are driven by the same long-range potentials.

We will present in the {    next subsection} the results for these particular S-waves  and
devote the last  one to describe some low-energy  properties of  higher angular momentum states.

\begin{table*}[thb!]
\begin{tabular}{l rr  c rr     l   rr }   \hline
                  &  \hspace{0.cm}Friar+&Dipole     &&      Kelly+            &\!\!Kelly         &&   Atac+                  &Kelly        \\\hline
$^1$S$_0$&   $a_0$\;\;\;\;\;\;\;      & $r_0$\;\;\;&&   $a_0$\;\;\;\;        & $r_0$\;\;\;\;  &&   $a_0$\;\;\;\;         & $r_0$\;\;\;\;         \\\hline
$e^-n$       &\;  2.926\,10$^{-3}$   & -149       && 2.920\,10$^{-3}$  & -186             &&  2.920\,10$^{-3}$  &    -186  \\
$\mu^- n$  &\;  2.501\,10$^{-3}$   &  -170      &&  2.497\,10$^{-3}$ & -215             &&  2.501\,10$^{-3}$  &  -215  \\
$\tau^- n$  &\;  1.574\,10$^{-4}$   &  4814     &&  1.623\,10$^{-4}$  &  -2145         &&  1.849\,10$^{-4}$ &  -276 \\\hline
$e^+n$      &\;  -2.949\,10$^{-3}$  & 150        && -2.943\,10$^{-3}$ &  186            &&  -2.943\,10$^{-3}$   & 186  \\
$\mu^+n$  &\;   -2.518\,10$^{-3}$ &  171       &&  -2.514\,10$^{-3}$ &  217          &&  -2.517\,10$^{-3}$ &  216 \\
$\tau^+n$  &\;  -1.577\,10$^{-4}$  & -4802\;   && -1.625\,10$^{-4}$ &  2142          &&  -1.851\,10$^{-4}$ &  276 \\
\end{tabular}
\caption{Low energy $ln$ parameters (in fm) in the $^1$S$_0$ state   {obtained
with different choices for the electric (first name heading each column) and magnetic (second name) $n$ form factors:
Friar (\ref{GEn_F}), Dipole (\ref{GMn_Dipole}), Kelly  \ref{GEn_Kelly})  and Atac et al. \cite{Atac_Nature_Com_2021} used to compute the potential}.}\label{Tab_LEP_1S0}
\end{table*}

\subsection{Low energy parameters for S-waves}\label{SSect_S}

We will start with the  coherent and incoherent LEPs for the $^1$S$_0$  and $^3$SD$_1$ states, for which they are  well defined. 
They will be completed by the low-energy phase shifts  and cross sections and compared to some experimental results
obtained in the low energy $n$ scattering on atomic systems.

\subsubsection{$^1$S$_0$}

For the S-wave singlet state ($^1$S$_0$) one has $\vec\sigma_l\cdot\vec\sigma_n$=-3, $\hat{S}_{12}=0$ and $\vec{L}\cdot\vec{s}_n$=0. 
The $ln$  potential is given in \eqref{V1S0}
As seen in Fig.~\ref{Fig_Vln_FD}, this potential is globally repulsive for all leptons and attractive for antileptons.
 
The  corresponding LEPs  are given in Table~\ref{Tab_LEP_1S0} in fm units. 
The different columns correspond to different choices of $G_E$ and $G_M$: Friar (\ref{GEn_F})+Dipole(\ref{GMn_Dipole}),  Kelly(\ref{GEn_Kelly})+Kelly(\ref{GMn_Kelly}) and 
Atac\cite{Atac_Nature_Com_2021}+Kelly combination of form factors. 
The upper half part of the table corresponds to lepton-neutron ($ln$) and the lower part to antilepton-neutron ($\bar{l}n$).
Several comments are in order:
\begin{itemize}
\item For $e$ and $\tau$ there is a nice stability in the predictions for the scattering length among the different $n$ form factor parametrisations.
This is due to the fact  that this quantity is essentially dominated by $V_S$, which is very similar in the three parametrisations.
For the $\tau$ lepton, the two components of the potential, $V_C$ and $V_S$, become comparable and the scattering length is
 sensitive to small differences in the $n$ charge and magnetic form factors.

The effective ranges, on the contrary, show clear discrepancies varying from  20\% in the $en$ and $\tau n$ cases
to more than a factor 10 in $\tau n$ (including sign).

\item For the $en$ case, the potential is dominated by $V_S$,  whose contribution, affected by  a factor -3, is strongly repulsive ($\sim 5\,$GeV).
However when the lepton mass increases, the repulsive $V_S$ term decreases (as  $m_e/\mu_l$ ) and  can  be compensated by the attractive $V_C$.
This is manifested by the decreasing value of the, still repulsive, scattering length  $a_{ln}$ in the upper part of Table~\ref{Tab_LEP_1S0}, which in the $\tau$ case is close to zero.
By  artificially increasing the lepton mass,  $a_{0}$  will become negative at $m_l\approx 1.18\, m_{\tau}$.

\item  If the problem was fully perturbative, that  is $T=V$ (where T is the T-matrix obeying the Lipmann-Schwinger equation), one should have $a_0({ln})+a_0({\bar l n})=0$.
As one can see from Table~\ref{Tab_LEP_1S0} by comparing the upper and lower half parts, this condition is quite accurately fulfilled, for $a_0$ as well as for $r_0$. 
In fact, the value  $s=a_0({ln})+a_0({\bar l n})$ {    constitutes a measurement of the non-perturbative effects, mainly due to two-photon exchange contributions:}
\begin{eqnarray*}
 &&T^{ln}+ T^{\bar ln}=    [V+ VG_0V + \ldots]  \\ &&+   [ (-V) + (-V)G_0(-V) + \ldots   ] =   2VG_0V + \ldots
\end{eqnarray*}
For the $e$ and $\mu$,  $s\approx 2\;10^{-5}$ fm, that is about 1\%, and for $\tau$ one order of magnitude smaller.   

\item In the limit of an infinitely heavy lepton, the potential is  given by the Coulomb term and the reduced mass $\mu_{ln}=m_n$.

\item The most favorable situation to obtain a $ln$ bound state concerns this channel, not for the $en$ case since it is repulsive, but for the positron  $e^+ n$
and anti-muon cases  for which $a\approx-3. 10^{-3}$ fm.
However, the very small values of these scattering lengths tell us that these systems are still very far from a possible bound state.
Its very existence would require changing the sign of $a_0$ after crossing a singularity.
 It can have some interest to see how far we are from an eventual binding and give no
place for eventual further speculations \cite{Grant_Cobble_PRL23_1969,Schlitt_PRB2_1970,McGuire_AJP40_1972,Schlitt_Letter_1973}. 
To this aim we have introduced an enhancement factor $\eta$ in front of the $V^{ln}_{1S0}$ potential
and determined the critical value of $\eta$ where $a_0\to +\infty$, indicating that a zero-energy bound state starts to appear. 
The result is $\eta_c=231$ for $e^+n$ and
$\eta_c=266$ for $\mu^+n$,  far beyond any reasonable uncertainty in the constructed potential.
\end{itemize}


\begin{table*}[htb!]
\begin{tabular}{l rr  c rr     l   rr }   \hline
                  &     Friar+Dipole         &               &&      Kelly+Kelly      &                     &&   Atac+Kelly           &     \\\hline
$^3$S$_1$&   $a_0$\;\;\;\;\;\;\;      & $r_0$\;\;\;&&   $a_0$\;\;\;\;        & $r_0$\;\;\;\;  &&   $a_0$\;\;\;\;           & $r_0$\;\;\;\;         \\\hline
$e^-n$       &\; -0.981\,10$^{-3}$   &  448        && -0.979\,10$^{-3}$ & 559             &&  -0.979\,10$^{-3}$   &   559  \\
$\mu^- n$  &\; -1.015\,10$^{-3}$   &  462        && -1.012\,10$^{-3}$ & 546             &&  -1.009\,10$^{-3}$   &   557  \\
$\tau^- n$  &\; -1.200\,10$^{-3}$   &  498        && -1.192\,10$^{-3}$ & 482             &&  -1.117\,10$^{-3}$   &    535 \\\hline
$e^+n$      &\;  0.979\,10$^{-3}$  & -448        &&  0.977\,10$^{-3}$ & -559             &&  0.977\,10$^{-3}$   &   -559   \\
$\mu^+n$  &\;  1.012\,10$^{-3}$ &  -461        &&  1.010\,10$^{-3}$ & -545             &&  1.006\,10$^{-3}$   &   -556 \\
$\tau^+n$  &\;  1.200\,10$^{-3}$  & -497        &&  1.189\,10$^{-3}$  & -481            &&  1.117\,10$^{-3}$   &   -535 \\
\end{tabular}
\caption{Low energy $ln$ parameters (in fm) in the $^3$S$_1$ state,   {with the same conventions as in Table. \ref{Tab_LEP_1S0}}.}\label{Tab_LEP_3S1}
\end{table*}

\subsubsection{$^3$S$_1$-$^3$D$_1$}

The  $^3$S$_1$-$^3$D$_1$ state  is a coupled channel  with the potential matrix
\[ V^{ln}_{^3S_1-^3D_1}=  \begin{pmatrix} V_C + V_{S} &  2\sqrt2 V_T  \cr   2\sqrt 2 V_T &    V_C + V_S  -2 V_T-{3\over2} V_{LS} \end{pmatrix} \, . \]
However, for this particular state, the diagonal tensor term is zero in the $^3$S$_1$ channel and the coupling to the $^3$D$_1$ channel is small, as its can be
seen from Fig.~\ref{Fig_VMMen}.
As a very good approximation we will first consider the $^3$S$_1$ channel alone: 
\[  V^{ln}_{^3S_1}(x) = V_C(x) + V_{S}(x) \, , \]
in which both components are attractive, giving rise to the unique $ln$ attractive channel, as seen in Fig.~\ref{Fig_Vln_FD}.
The corresponding LEP parameters are displayed in Table~\ref{Tab_LEP_3S1} for the same combinations of $n$ form factors as in Table \ref{Tab_LEP_1S0}.

When compared to the $^1$S$_0$ state one first remarks a much higher stability in the predictions of different form factors, including
the $\tau$ lepton and the effective range parameter $r_0$. This is  due to the absence of any compensation
between the Coulomb ($V_C$) and  magnetic ($V_S$) terms, which are both attractive.

\begin{figure}[htbp]
\centering\includegraphics[width=9cm]{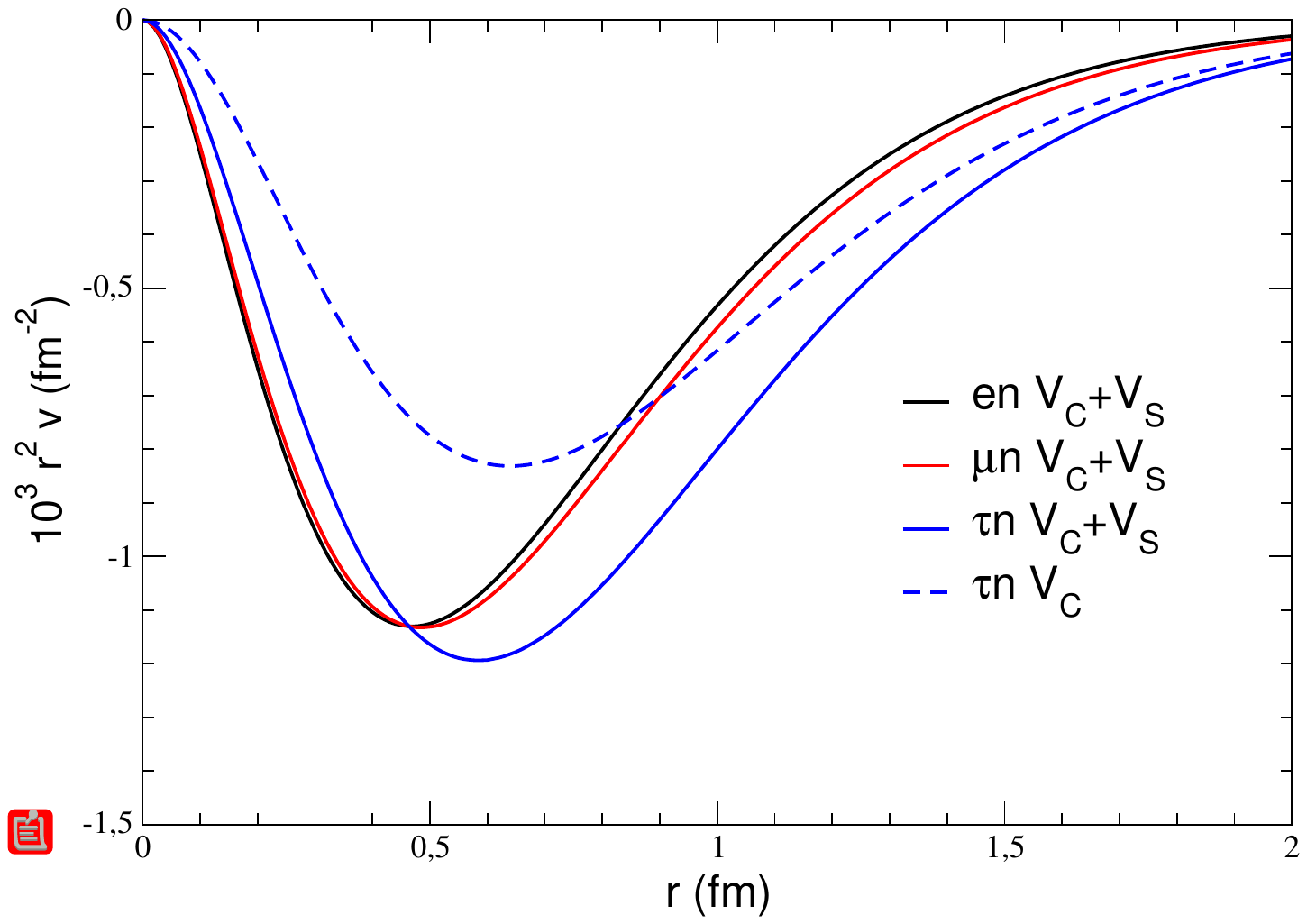}
\caption{Reduced $^3$S$_1$ potentials   {(\ref{Vef})} multiplied by $r^2$ for the three different $ln$ systems depicted by solid lines.  The dashed line is the reduced Coulomb potential for the $\tau n$ system.}\label{Fig_Reduced_V3S1}
\end{figure}

One can remark also a  kind of  flavour independence of the $ln$ scattering lengths: they vary about 20\% while the lepton masses vary over three orders of magnitude.
This is the combined consequence of, on one hand, a purely attractive channel (no cancellations between $V_C$ and $V_S$) and on the other hand, the fact that the reduced spin-spin potential scales as:
\begin{equation}\label{Vef}
 v_{S}\equiv{2\mu\over \hbar^2}V_S\sim {m_lm_n\over m_l+m_n} {1\over m_l m_n} \sim  {1\over m_l+m_n}\, .
 \end{equation}
For the $en$ and $\mu n$ systems, $v_S$ is the dominant contribution of the total potential $v^{ln}$, 
while for $\tau n$, $v_S$  is suppressed by a factor of $\sim m_n/(m_\tau+m_n)\approx 1/2$ with respect to $en$ and $\mu n$ and becomes comparable and even smaller 
than the reduced Coulomb potential $v_C$. The final reduced potential $v=v_C+v_S$, in the region of interest to determine the scattering length,
\begin{equation}\label{a0_B}
 a_0^B= \int_0^{\infty} dr\; r^2\; v(r) 
 \end{equation}
turns to be roughly independent of the lepton mass.
We have illustrated this fact by plotting  in Fig.~\ref{Fig_Reduced_V3S1} the integrand of eq. (\ref{a0_B}) for the different leptons as well as the purely Coulomb potential. 

Notice also that the non-perturbative effects are one order of magnitude smaller than for $^1$S$_0$, with $s\approx 10^{-6}$\,fm for $e$ and $\mu$.
The coupling to the $^3$D$_1$  channel by the small tensor force $V_T$ does not modify sizeably the value value of the $^3$S$_1$ scattering lengths
given in Table~\ref{Tab_LEP_3S1}.

Concerning the possibility of an eventual bound $en$ state in this channel, the critical enhancement factor -- defined in the previous subsection --  
is $\eta_c=690$  for $en$, roughly a factor 3 larger than for  $^1$S$_0$ state, the same factor that exists between the respective potentials.

\begin{figure*}[htbp]
\centering\includegraphics[width=5.5cm]{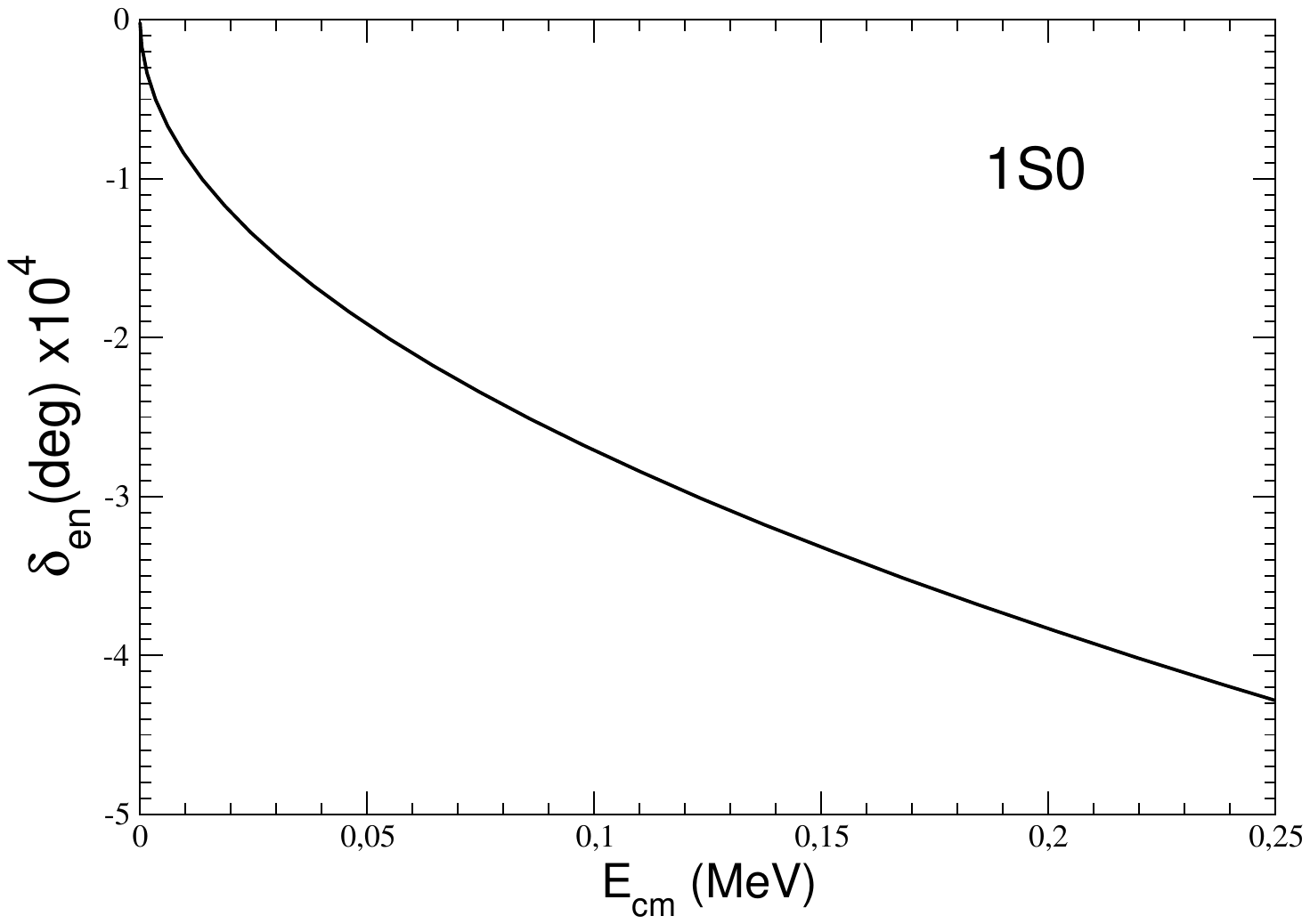}
\centering\includegraphics[width=5.5cm]{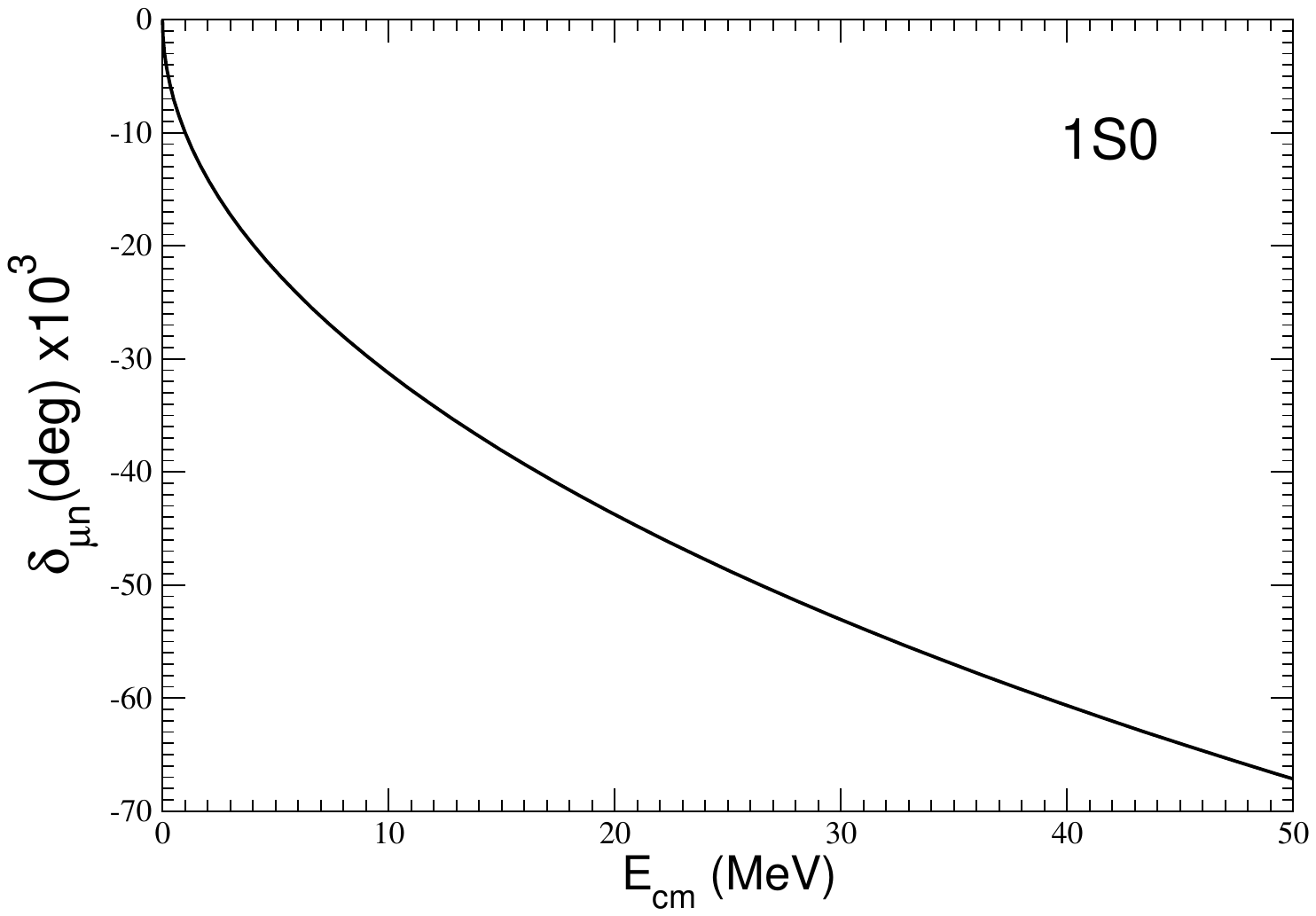}
\centering\includegraphics[width=5.5cm]{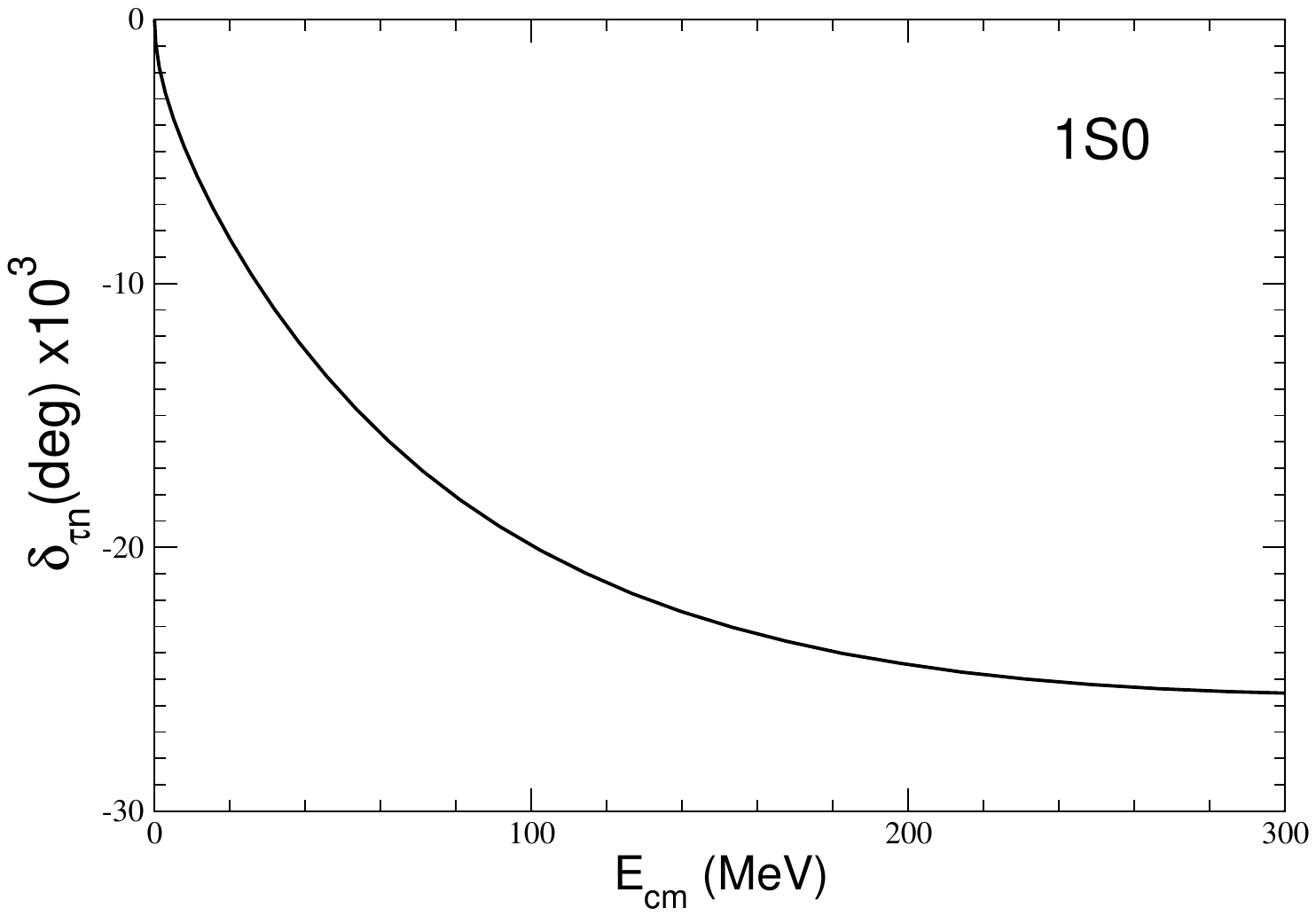}

\centering\includegraphics[width=5.5cm]{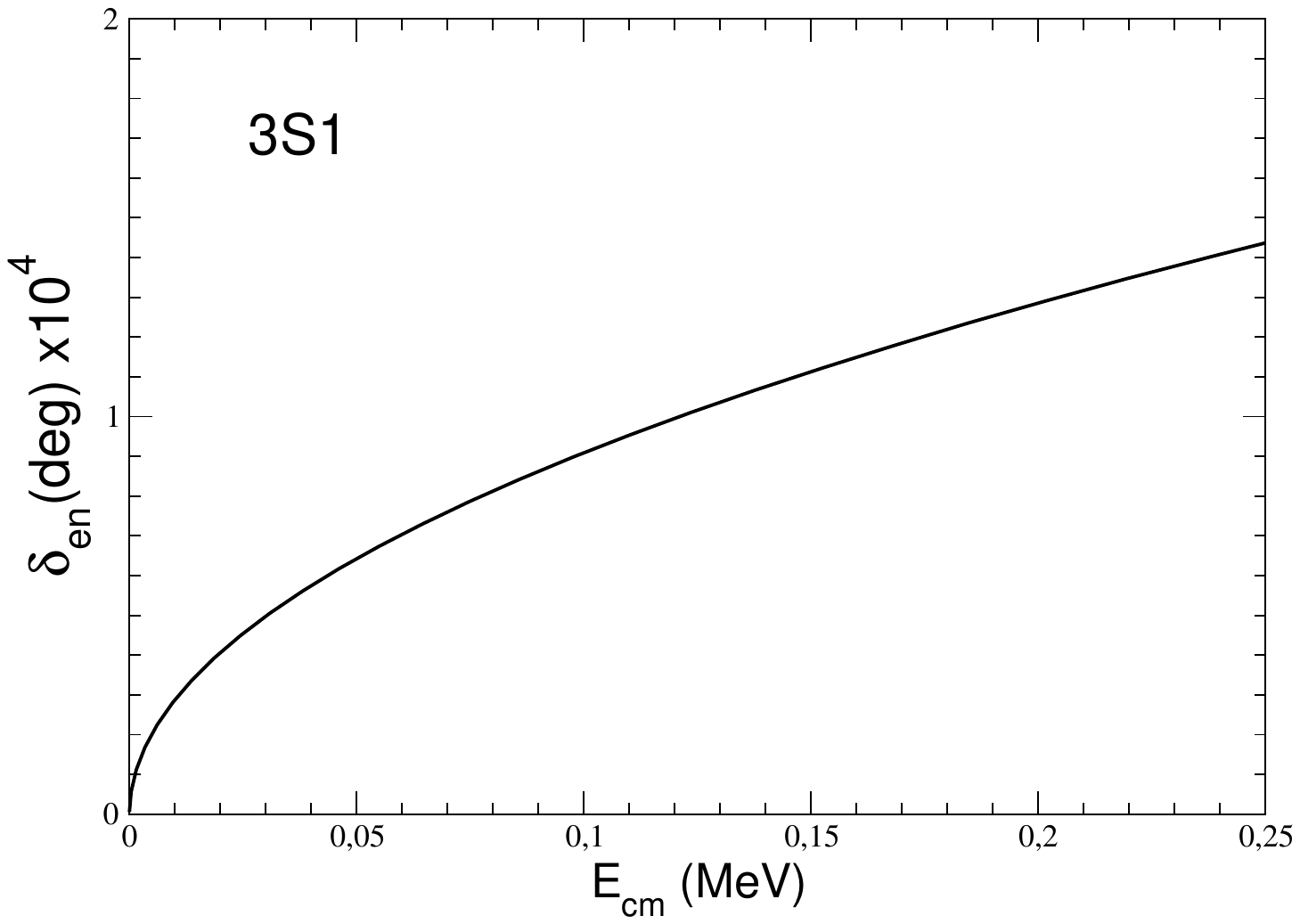}
\centering\includegraphics[width=5.5cm]{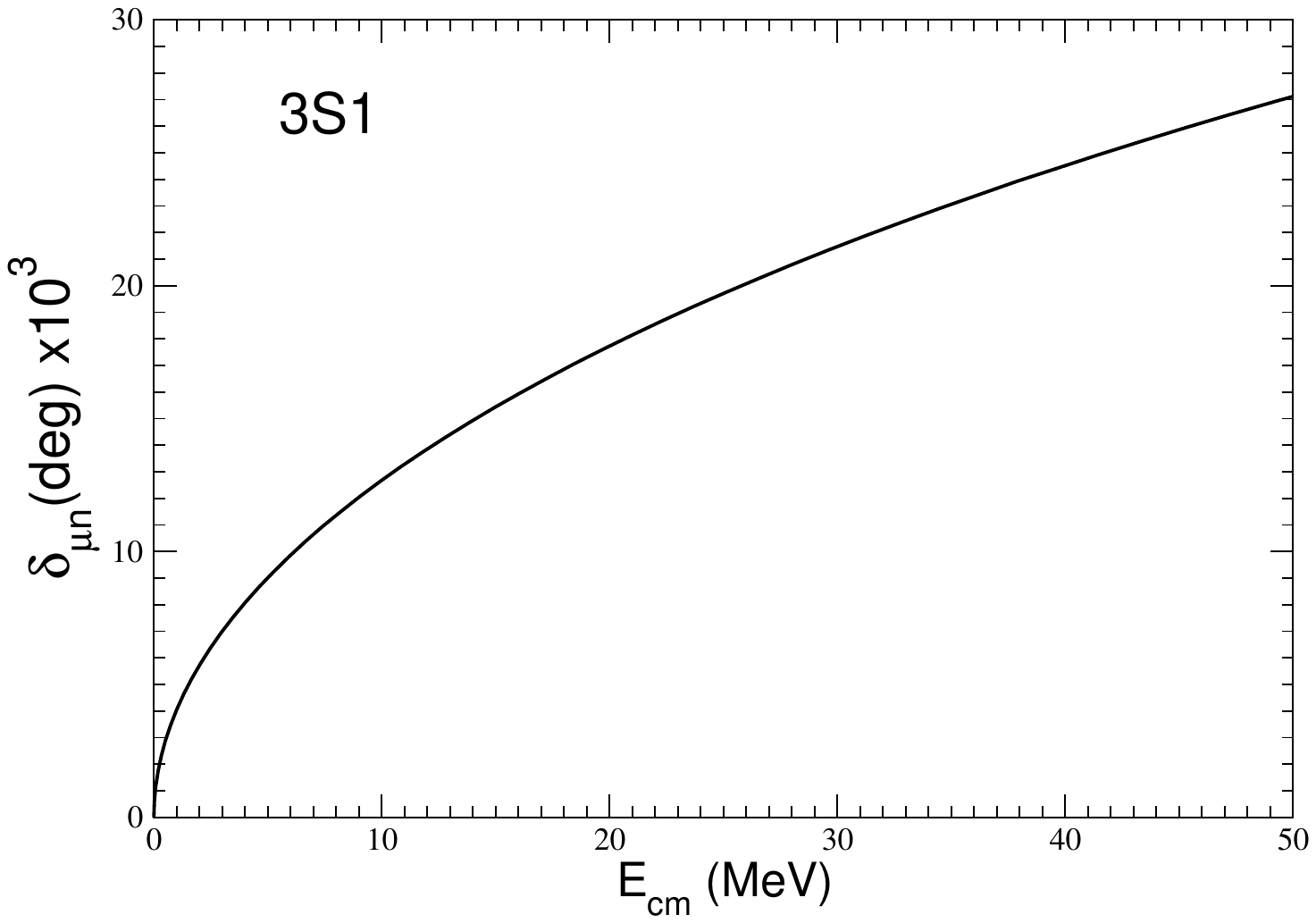}
\centering\includegraphics[width=5.5cm]{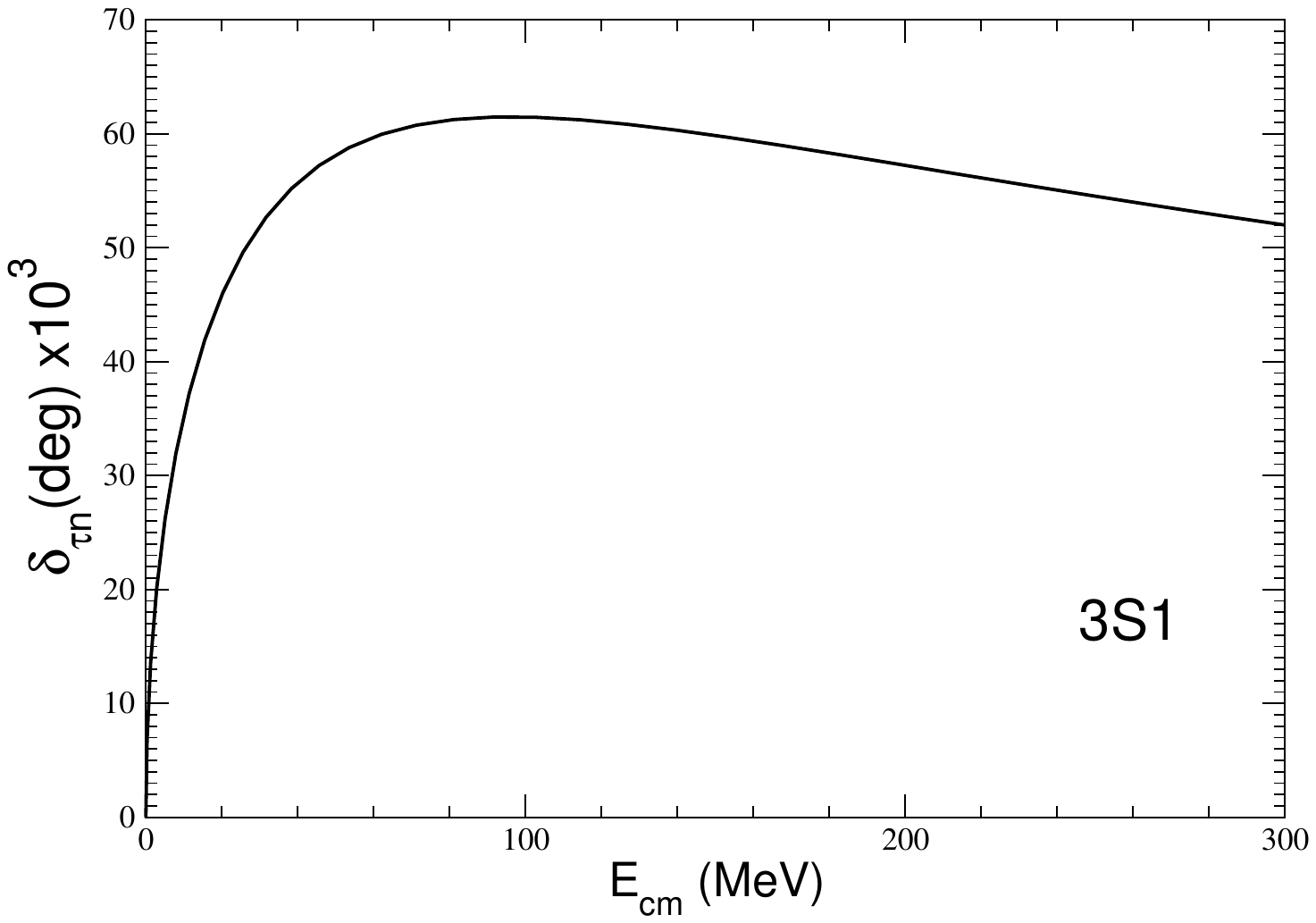}
\caption{$ln$ S-wave phase shifts (in degrees) as a function of cm energy $E_{cm}$.}\label{Fig_delta_ln}
\end{figure*} 

\subsubsection{S-wave phase shifts, cross sections and coherent scattering lengths}\label{SSect_delta_sigma}

The $ln$ phase shifts have been computed   by solving the Schr\"odinger equation up   {to a center of mass} energy $E_{cm}$ half of the $ln$ reduced mass $\mu_{ln}$, i.e. up to
center of mass momentum  $k_{max}={\mu_{ln}/\hbar c}$.
This defines our kinematical constraint; beyond, a relativistic dynamics would be required.
In this kinematical domain, the computed phases are accurately reproduced  by the effective range expansion 
 \begin{multline}
 \label{ERE}
    k\cot\delta_0(k)= -{1\over a_0} + {1\over 2} r_0 k^2 \\ \Longrightarrow\quad \delta_0(k)= -k\,a_0\left[  1 + {1\over2} r_0\,a_0 \,k^2 + \ldots \right]\, .     
 \end{multline}
with parameters given in Tables~\ref{Tab_LEP_1S0} and \ref{Tab_LEP_3S1}.

They are represented in Fig.~\ref{Fig_delta_ln} as a function  of $E_{cm}$.
As one can see,  all phase shifts are very small in the considered kinematical region. 
As expected, the perturbative treatment gives accurate result,  up to a degree that we have discussed in the previous section. 

Obtaining a departure from the linear behaviour at the origin given by $\delta_0(k)=-k\,a_0$ in (\ref{ERE}), would require  $r_0\,a_0\,k^2\sim 1$.
As one can see from Tables~\ref{Tab_LEP_1S0} and  \ref{Tab_LEP_3S1}, the product $r_0\,a_0$ takes, for both S states and all considered leptons,  similar values $\sim 0.4-0.8$ fm$^2$.  
Thus, the effective range manifests only above $k\sim 1\,$fm$^{-1}$, which is -- between our kinematical constraint --  realized only for the $\tau n$  ($k_{max}$=3.11 fm$^{-1}$), and to a less extent for $\mu n$ ($k_{max}$=0.48 fm$^{-1}$).
In the $en$ case the phase shifts are accurately given by $\delta_0(k)=-a_0 k$.
 
The $ln$ {    total S-wave cross section} takes the form
\[  \sigma_{ln}(k) = {1\over 4}  \sigma_s(k) + {3\over 4}  \sigma_t(k)   \quad \text{with}\quad    \sigma_{i=s,t}= 4 \pi { \sin^2\delta_i(k)\over k^2} \,, \]
where  the index $s$ denotes the singlet $^1$S$_0$ state, $t$ the triplet $^3$S$_1$.
The zero-energy limit is given by
\[ \sigma_{ln}(0)= \pi ( \mid a_s\mid^2 + 3 \mid a_t\mid^2 )\, , \]
 and provides  similar values for the  three considered leptons: $\sigma_{en}(0)$=\,0.358\,$\mu b$,  $\sigma_{\mu n}(0)$=\,0.292\,$\mu b$,
  $\sigma_{\tau n}(0)$=\,0.136\,$\mu b$~\footnote{1\,$\mu b=10^{-4}$\,fm$^2$}.

\bigskip
Before concluding this section it is worth  considering the $ln$ {    coherent scattering length},  defined as
\begin{equation}\label{ac}
 a_c= {a_s + 3\; a_t\over 4} \,.
\end{equation} 

By inserting in (\ref{ac}) the results of Tables~\ref{Tab_LEP_1S0} and  \ref{Tab_LEP_3S1}, one gets 
the  $a_c$  values  displayed in  the upper half part of Table~\ref{Tab_ac} (in fm). 
For the $en$ and  $\mu n$ cases,  there is a remarkable stability with respect the different choices of form factors
but for  $\tau n$ they can differ by up to 50\%.

{    Notice that, in the Born approximation, i.e. $T\equiv V$, the coherent scattering length (\ref{ac})
would be entirely given by the spin-independent Coulomb potential $V_C$}.
Indeed, in this case the singlet ($^1$S$_0$) and triplet ($^3$S$_1$) contributions to $a_c$  coming from the spin-spin magnetic term $V_S$  would exactly compensate each other, due to
the $(\vec\sigma_l\cdot\vec\sigma_n)$ term, and any non zero value of  $a_c$ would entirely come from $V_C$.

We can check this fact by switching off the magnetic term $V_S$ in the potentials and obtain in this way
 the "pure Coulomb" coherent scattering length, denoted by $a_c^C$. The result is given in the lower half part of Table~\ref{Tab_ac}. 
For the $\mu n$  case,  $a^C_c$  and $a_c$ are indeed practically identical, and for $\tau n$ both quantities are very close.
However {     in the $en$ case, the value of  the coherent scattering length $a_c$ is one order of magnitude larger than  what one could expect  from the Coulomb potential
alone ($a^C_c$).}
It follows from that the value of the "in flight" $en$ coherent scattering length is dominated by, and measures, the non-perturbative effects in the $en$ scattering process.
The dynamical {    reason for this difference is the huge  value of the spin-spin potential in the $en$ case}.

\begin{table}[thb!] 
\begin{tabular}{l       c                         c           c                              c      c                    r } \hline
                       &    Friar+Dipole         & &      Kelly+Kelly              & &   Atac+Kelly                 &   \\\hline
                       &   $a_c$                    &&   $a_c$                          & &  $a_c$                         &        \\\hline
$e^-n$\;\;\;\;    & -4.50\;10$^{-6}$       & &  \;\;\;\;-4.42\;10$^{-6}$ & & \;\;\;\;-4.43\;10$^{-6}$  &                    \\
$\mu^-n$\;\;\;\;& -1.36\,10$^{-4}$      & &   \;\;\;\;-1.35\;10$^{-4}$  & & \;\;\;\;-1.32\;10$^{-4}$   &                    \\
$\tau^-n$\;\;\;\;& -5.07\,10$^{-4}$      & &  \;\;\;-4.88\;10$^{-4}$     & & \;\;\;-3.76\;10$^{-4}$      &   \\\hline
                       &   $a^C_c$               &&   $a^C_c$                       & &  $a^C_c$                      &       \\\hline
$e^-n$\;\;\;\;    &-7.14\;10$^{-7}$      & & \;\;\;   -7.08 \;10$^{-7}$   & & \;\;\;-6.90 \;10$^{-7}$    &                    \\
$\mu^-n$\;\;\;\;&-1.33\;10$^{-4}$      & & \;\;\;  -1.32 \;10$^{-4}$    & & \;\;\;-1.28 \;10$^{-4}$    &                    \\
$\tau^-n$\;\;\;\;&-8.60\;10$^{-4}$       & & \;\;\;   -8.53\;10$^{-4}$    & & \;\;\; -8.31\;10$^{-4}$    &                    \\ 
$e^+n$\;\;\;\;    &\;7.14\;10$^{-7}$      & & \;\;\;\;   7.08 \;10$^{-7}$   & & \;\;\;\;6.90\;10$^{-7}$    &    \\\hline
\end{tabular}
\caption{Coherent $ln$ scattering lengths $a_c$ and the value $a_c^C$ produced by the Coulomb potential $V_C$  only (in fm units).}\label{Tab_ac}
\end{table}

\begin{figure}[htbp]
\centering\includegraphics[width=9.cm]{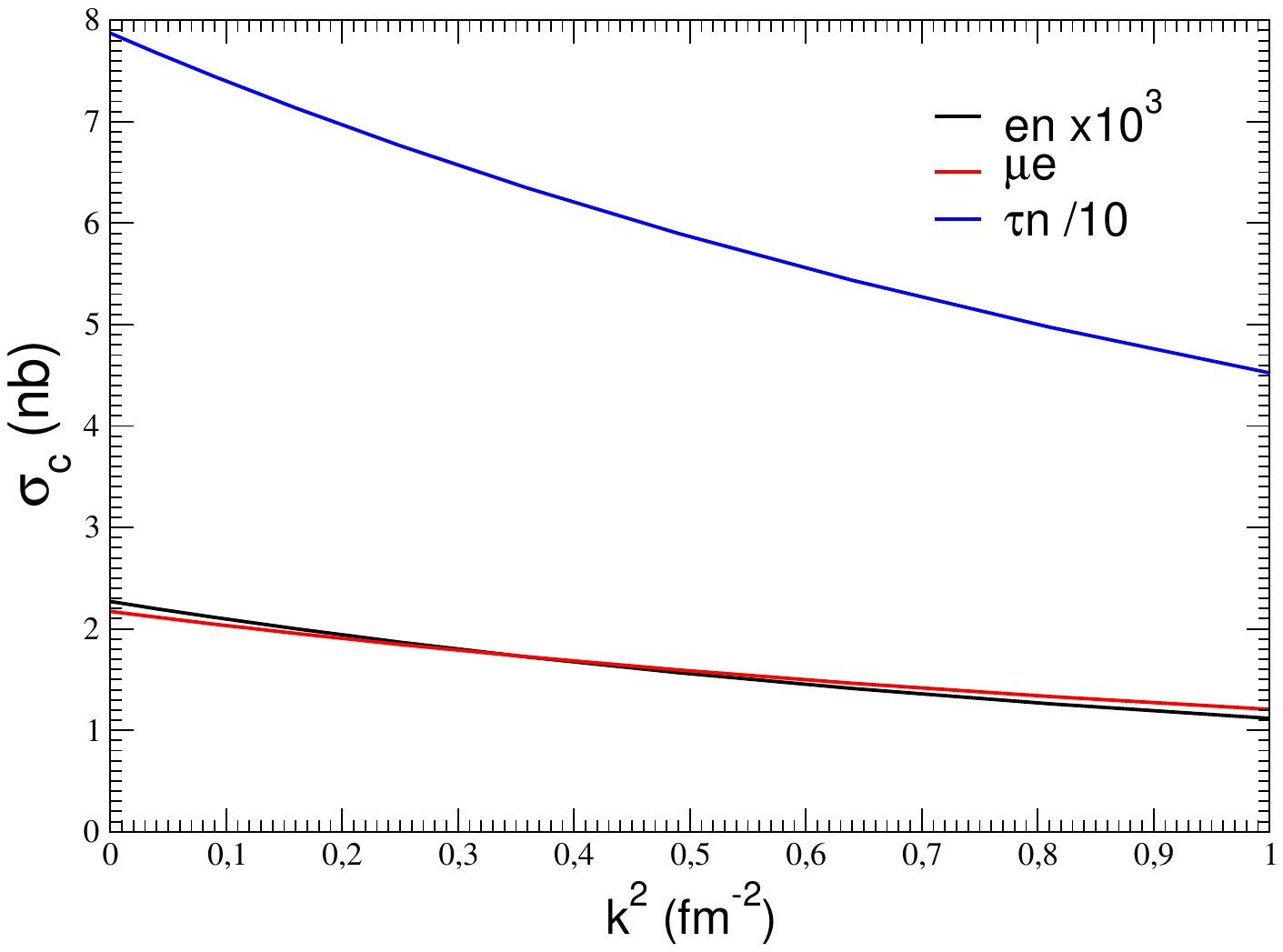}
\caption{Coherent $ln$ scattering cross section (in nb).}\label{Fig_sigma_c}
\end{figure}

Finally the {    coherent  scattering cross sections},  given by
\[   \sigma_c= 4\pi\; \Big|  {1\over4} \left[    f_s(k)  + 3 f_t(k)  \right]\Big| ^2 \,,\]
are represented in Fig. \ref{Fig_sigma_c} as a function of $k^2$.  They correspond to Atac+Kelly $n$ form factors.
The zero-energy  coherent cross section  is  $ \sigma_c$=0.0023 $nb$  for  $en$,   $ \sigma_c$=2.2 $nb$ for $\mu n$ and 
 $ \sigma_c$=79 $nb$ for $\tau n$, that is 3-5 orders of magnitude smaller  than the incoherent cross sections (1 $nb$=10$^{-3}$ $\mu b$=10$^{-7}$ fm$^2$).

To our knowledge, there has been no any measurement   {of either  the coherent
or the incoherent}  $ln$ cross sections, although it was used in some experiments as a fit parameter
for determining the n-"electron-bound-in-a-heavy-atom" coherent scattering length, which is the topic of the next section.

\subsubsection{$n$ scattering on an "$e$-bound-to-atom"}\label{Sect_e*n}

 \bigskip 
A very different situation occurs if {    one assumes}, as was done in the Foldy seminal paper~\cite{Foldy_RMP30_1958}	and subsequent reviews 
on this topic~\cite{Sears_PR141_1986,Abele_PPNP_60_2008}, 
that {    a very low-energy $n$} (termal energies) {    scatters coherently on a single electron, bound in a heavy atom,   {which}  recoils as a whole}. 
In this case {    the electron can be considered as being infinitely heavy}  and the reduced mass of the $ne$ 
system is equal to the neutron mass: $\mu_{en}\equiv m_n$. 

One rather talks about  the   { coherent scattering length of a  {    $n$ colliding with an ''electron-bound-to-atom"}}, abusively shortened into  $ne$ coherent scattering length, and traditionally denoted $b_{ne}$. 
In order to distinguish this process  from the "on-flight" $ne$ one described in the previous section, we will use for the former case the notation $ne^*$
and  the corresponding coherent scattering length by $a_c(ne^*)\equiv b_ne$.

{    In this approach, the magnetic interaction is disregarded and  $V^{ne^*}$ is simply given by the Coulomb term $V_C$ -- i.e. the $n$ charge form factor  $G_E$ -- 
which entirely determines the value of $b_{ne}\equiv a_c(ne^*)$}.
\footnote{The word coherent is also abusive in this respect, since it is associated to a spin-independent potential $V_C$, although strictly speaking, the relation~(\ref{ac}) is trivially fulfilled.}

One obtains in this way the $b_{ne}$ values displayed in Table~\ref{Tab_LEP_bne}, together with the corresponding effective ranges.
Notice  a three order of  magnitude enhancement, of purely kinematical origin, with respect the $en$  on-line coherent  scattering lengths given in Table \ref{Tab_ac}.

The $b_{ne}$ values presented in Table~\ref{Tab_LEP_bne} are  in close agreement with the experimental value $b_{ne}=-1.32\pm0.03\,$fm from~\cite{Koester_PRC51_1995}.
\begin{table*}[thb!]
\begin{tabular}{  r                             r                             r                              r                            r                                  r } \hline
  Friar                    &                    &\;\;\;\; Kelly                  &                              &   Atac                           &   \\\hline
$a_0(ne^*)$\;\;\;\;  & $r_0(ne^*)$& \;\;\;  $a_0(ne^*)$      & $\;r_0(ne^*)$\;\;\    &  $a_0(ne)$\;\;\;\;\;\;\;     &   $\;\;\;r_0(ne^*)$\;\;       \\\hline
  -1.32\,10$^{-3}$  &  501\;\;\;      &\;\;\;-1.31\,10$^{-3}$\;\;& 449           \;\;\;\;    & \;\;\;\;-1.27\,10$^{-3}$\;\;&  \;\;\;\;518    \;\;\;\               \\ \hline
\end{tabular}
\caption{$n$-{    electron-bound-to-atom} ($ne^*$)  coherent scattering lengths $a_0(ne^*)\equiv b_{ne}$ (in fm) produced by $V_C$ only and with different charge form factors.}\label{Tab_LEP_bne}
\end{table*}
It is worth noticing that the value of $b_{ne}$ -- entirely determined by $G_E$ -- is strongly dominated  
by the the so-called  Foldy term \cite{Foldy_RMP30_1958,GR_PLB459_1999,BC_PRC60_1999,VAK_NPA699_2002}, that is the 
contribution due to the $F_2$ Dirac form factor in the standard decomposition 
of $G_E$ \cite{BD_RQM_1964}:
\[ G_E(q^2)= F_1 (q^2)+ {q^2\over 4m_p^2}\kappa \; F_2(q^2)\,.\]

By considering  the $ln$ Coulomb interaction in momentum space
$$V_C\equiv - {\alpha(\hbar c)\over 2\pi^2} \;{G_E(q^2)\over q^2}$$
 and applying expansion (\ref{Gq2_exp}) and (\ref{r2k}) to $G_E$,  one  obtains
at the lowest order in $q^2$, the Born  approximation of the $en$ scattering length  displayed in Table~\ref{Tab_LEP_bne} in terms of the  moments of the $n$ radial charge density
\cite{Abele_PPNP_60_2008} 
 \begin{eqnarray}\label{a0_Born}
a_0^B(e^*n) &=& {(m_n c^2)\alpha\over 3(\hbar c)}  \langle r^2\rangle_n\, , 
\end{eqnarray}  
We have shown that the next order in $q^2$ provides the effective range values
 \begin{eqnarray}\label{r0_Born}
r^B_0(e^*n)         &=& -{1 \over 5 \,a_0^B(ne^*)} \;  {\langle r^4\rangle_n \over \langle r^2\rangle_n} \,.
\end{eqnarray}  
Due to the perturbative character of the interaction, these relations provide quite accurate results and give some light to the large values of the effective ranges obtained.
The former expressions can  be generalized to the  incoherent LEPS  from Tables~\ref{Tab_LEP_1S0} and \ref{Tab_LEP_3S1}, provided one properly includes the 
contribution due to the magnetic form factor $G_M$.

\bigskip
Last but not least, we would like to emphasize that, if one takes into account the full magnetic interaction (even a fixed electron keeps its magnetic moment) the results change dramatically.
The values of the coherent  ($a^*_c$) and  incoherent  ($a_{s,t}^*$) scattering lengths are given in Table~\ref{Tab_LEP_eA} for the different choices of $n$ densities.
When compared to the results of Tables~\ref{Tab_LEP_1S0} and \ref{Tab_LEP_3S1} one can see a 3 orders of magnitude enhancement due to the kinematical factor.
One can remark also a positive sign for the triplet scattering length, whose potential
is purely attractive. This indicates the formation of a n-$e^*$ bound state in this particular channel. 
Its binding energy is $B\approx 110$\,MeV and the  rms radius $R=\sqrt{\langle r^2\rangle }$=0.55\,fm.
This state  corresponds to  a pole in the $n-e^*$ scattering amplitude in the physical energy sheet, although
the experimental pertinence of such a result is not clear.

\begin{table}[thb!]
\begin{tabular}{l      c                    c                             c                                 c                               c               c   } \hline
                     &    Friar+Dipole &                  &      Kelly-Kelly                &                             &   Atac-Kelly  &          \\\hline
$a_s^*$        &  0.843            &                   &  0.905                           &                             &       0.905                              &                     \\
$a_t^*$         &  0.611            &                   &  0.567                           &                             &       0.567                     &                     \\
$a_c^*$        &  0.669           &                   &  0.652                            &                             &       0.652                                &    \\
\hline
\end{tabular}
\caption{$n$-electron-bound-to-atom  incoherent  ($a_s^*$ and $a_t^*$) and coherent ($a^*_c$) scattering lengths (in fm) produced by the full $V^{ln}$ interaction.}\label{Tab_LEP_eA}
\end{table}

\subsection{Higher partial waves}\label{SSect_P}

All the $ln$ states  with non-zero angular momenta ($J^{\pi}=0^-,1^-,2,^-,2^+,3^+,3^-,\ldots$)  involve  $\mid SLJ\rangle$ coupled channels.
This coupling is  produced by $1/r^3$ long-range potentials which are due  to the tensor ($V_T$) and spin-orbit  ($V_{LS}$) terms in (\ref{Vln}).
Solving the coupled-channel scattering problem  with a long-range coupling between channels requires specific methods, like those developed in Refs \cite{VDLH_PRA97_2018,MVLH_HI239_2018}, 
and it is beyond the scope  of the present work.

However {    one can obtain an estimation} of  the scattering amplitude and cross section for non-zero angular momentum states {    by neglecting the coupling 
among the channels} and considering decoupled $\mid SLJ\rangle$ states. Still we will be faced with the non trivial problem of  scattering by a short-range plus asymptotically $1/r^3$ potential.
It is well known from the  early sixties \cite{SOMR_PRL5_1960,SOMR_JMP2_1961,GG_NC38_1965_I,GG_NC40_1965_II,Shakeshaft_JPB5_1972} 
that the $1/r^3$ asymptotic behaviour of the interaction precludes the existence of  low-energy parameters.

\begin{table}[ht!] 
\begin{tabular}{l       c                         c           c                              c      c                    r } \hline
                       &    Friar+Dipole         & &      Kelly+Kelly              & &   Atac+Kelly                 &   \\\hline
$^1$P$_1$     &   $a_1$    (fm$^3$) &&   $a_1$  (fm$^3$)          & &  $a_1$   (fm$^3$)          &        \\\hline
$e^-n$\;\;\;\;    &  2.15 \,10$^{-4}$     & &  \;\;\;\;2.67\;10$^{-4}$    & & \;\;\;\;2.67\;10$^{-4}$  &                    \\
$\mu^-n$\;\;\;\;&  1.79\,10$^{-4}$      & &   \;\;\;\;2.27\;10$^{-4}$   & & \;\;\;\;2.26\;10$^{-4}$   &                    \\
$\tau^-n$\;\;\;\;& -2.00\,10$^{-4}$      & &  \;\;\; 0.94\;10$^{-5}$     & & \;\;\;\;0.16\;10$^{-5}$      &   \\\hline
$^1$D$_2$     &   $a_2$  (fm$^5$)   &&   $a_2$     (fm$^5$)       & &  $a_2$   (fm$^5$)           &        \\\hline
$e^-n$\;\;\;\;    &  1.41 \;10$^{-5}$     & &  \;\;\;\;3.01\;10$^{-5}$    & & \;\;\;\;3.01\;10$^{-5}$  &                    \\
$\mu^-n$\;\;\;\;&  1.11\;10$^{-5}$      & &   \;\;\;\;2.60\;10$^{-5}$   & & \;\;\;\;2.57\;10$^{-5}$   &                    \\
$\tau^-n$\;\;\;  & -0.34\;10$^{-5}$      & &  \;\;\; 0.39\;10$^{-5}$     & & \;\;\;\;0.18\;10$^{-5}$      &   \\\hline
\end{tabular}
\caption{Singlet P- and D-wave $ln$  scattering "volumes" (the spin-orbit coupling to $^3$P$_1$ is neglected).}\label{Tab_a_1PD}
\end{table}

The simplest case is provided by the spin singlet states, only affected by the short-range $V_C$ and $V_S$, and  for which we can compute the LEPs.
The results for $^1$P$_1$ and $^1$D$_2$ scattering "volumes" are represented in Table \ref{Tab_a_1PD}. 
One can see  very small values of the corresponding scattering volumes with a net decreasing  as a function of $L$: one order of magnitude each L (10$^{-3}$ for $L=0$, 10$^{-4}$  for $L=1$, 10$^{-5}$ for L=2).
When compared to the S-wave results, one see  also a stronger dependence on the choice of $n$ density parametrizations: a sign inversion for $\tau$ is present in Friar-Dipole and absent in the other choices, due to the differences in the magnetic form factors.
Notice however that these results have only an informative character concerning the short-range part of the interaction, which
in its turn produces low-energy partial cross section  behaving as $\sigma_L(k)=\mid a_L\mid^2  k^{4L}$, and so vanishing at $k=0$.  
Notice also, that the spin-orbit coupling (see Table~\ref{Ops}) between he singlet states ($^1$L$_{J=L}$) and the corresponding natural parity triplet states ($^3$L$_{J=L}$)
could dramatically modify the zero energy scattering properties.

\bigskip
For all the other $L>0$  states, the $1/r^3$ behaviour of the interaction prevents a similar study.
However, it has been shown in recent works \cite{Gao_PRA59_1999,Muller_PRL110_2013}
that it is possible to obtain a simple expression for the zero energy cross section, which, contrary to what happens in the case of short-range interactions,
does not vanish  in the zero energy limit. 

\newcommand\tendsto{ \mathop{\longrightarrow} }

The key parameter is the asymptotic coefficient  $\beta_3$ of the (reduced)  long range interaction
\begin{equation}\label{beta3}
 \beta_3= {2\mu\over\hbar^2} \; C_3    \qquad {\rm with} \quad   C_3 = \lim_{r\rightarrow \infty}  r^3\; V(r)
\end{equation} 
It  has the dimensions of a length and, in our particular case, it depends on the  partial wave $\beta_3=\beta_3(L,S,J)$.
Since the central and  spin-spin terms in the the $V^{ln}$ potential are exponentially decreasing, $\beta_3$ has contributions coming
from the tensor  and from the spin-orbit potentials. They are obtained by multiplying the asymptotic constants of $V_T$ \eqref{VT_Ass} and $V_{LS}$ \eqref{VLS_Ass} -- 
which depends on the lepton flavour and on the $n$ form factor parametrisation --
 by  the corresponding matrix elements of $S_{12}$~\eqref{S12} and $\vec{L}\cdot \vec{s}_n$~\eqref{Lsn_SLJ_UPS}-\eqref{Lsn_SLJ_NPS}.

\begin{table*}[th!]
\begin{tabular}{l rr  c rr     l   rr }   \hline
                  &   \;\;\; $e^+n$          &                   &&\;\;  \;\;\;\;$\mu^+n$&                          &&   $\tau^+-n$           &     \\\hline
                  &   $\beta_3$\;\;\;\;\;\; &$\sigma_L(0)$&&\;\;$\beta_3$\;\;\; &$\sigma_L(0)$&&$\beta_3$\;\;\;\;\;\;   & \;\;\;$\sigma_L(0)$    \\ \hline
$^3$P$_0$&\;  1.76\,10$^{-2}$   &\;\;0.61\;      &&  1.70\,10$^{-2}$ &\;\;0.57\;&&  1.38\,10$^{-2}$   &  \;\;0.37\;\;       \\
$^3$P$_1$&\;  0.88\,10$^{-2}$   &\;\;0.45\;      &&  0.91\,10$^{-2}$ &\;\;0.49\;&&  1.07\,10$^{-2}$   &  \;\;0.68\;\;  \\
$^3$P$_2$&\; -5.28\,10$^{-3}$   &\;\;0.28\;      && -5.34\,10$^{-3}$ &\;\;0.28\;&& -5.67\,10$^{-3}$   &  \;\;0.32\;\;  \\ \hline
$^3$D$_1$&\;  2.06\,10$^{-2}$  &\;\;0.28\;       &&  2.03\,10$^{-2}$ &\;\;0.27\;&&  1.86\,10$^{-2}$   &  \;\;0.23\;\;   \\
$^3$D$_2$&\;  0.88\,10$^{-2}$  &\;\;0.08\;     &&  0.91\,10$^{-2}$ &\;\;0.09\;&&  1.07\,10$^{-3}$    &   \;\;0.12\;\;  \\
$^3$D$_3$&\; -1.09\,10$^{-2}$  &\;\;0.18\;     && -1.10\,10$^{-2}$ &\;\;0.18\;&& -1.15\,10$^{-3}$    &  \;\;0.20\;\;  \\ \hline
$^3$F$_2$&\;  2.58\,10$^{-2}$   &\;\;0.18\;       &&  2.56\,10$^{-2}$ &\;\;0.18\;  &&  2.43\,10$^{-2}$ &   \;\;0.16\;\;     \\
$^3$F$_3$&\;  0.88\,10$^{-2}$   &\;\;0.03\;     &&  0.91\,10$^{-2}$ &\;\;0.03\;&&  1.07\,10$^{-2}$     &   \;\;0.04\;\;  \\
$^3$F$_4$&\; -1.66\,10$^{-2}$   &\;\;0.14\;    && -1.67\,10$^{-2}$ &\;\;0.13\;  && -1.73\,10$^{-2}$   &   \;\;0.15\;\;  \\ \hline
\end{tabular}
\caption{Asymptotic coefficients $\beta_2$ (\ref{beta3})   (in fm) and zero-energy partial cross sections (\ref{sigma_LSJ})  (in $\mu$b) for the lowest non-zero angular momentum states.}\label{Tab_beta3_sigmaL}
\end{table*} 

It was shown  in Refs.~\cite{Gao_PRA59_1999,Muller_PRL110_2013}  that,
in the low-energy limit, the PW  phase shifts for $L>0$ are given by~\footnote{there is a minus sign in \cite{Gao_PRA59_1999}  due to different phase convention.}
\begin{equation}\label{tgd_LSJ}
 \tan\delta_{L,S,J}(k)= {1\over 2L(L+1)} k\beta_3(LSJ) + O(k^2)
 \end{equation}
which entirely depends on the asymptotic coefficient $\beta_3$ and it is independent of the short-range phase shifts. The scattering amplitude is, in this limit, given by
\[ f_{LSJ}(k) = {\tan\delta_L(k)\over k} + O(k) = {\beta_3(LSJ)\over 2L(L+1)}  + O(k)\]
and the partial cross section
\begin{multline}\label{sigma_LSJ}
\sigma_{LSJ}(k)= (2J+1) \pi  \mid f_{LSJ} \mid^2 \\ = (2J+1)   { \pi \beta_3^2  \over 4L^2 (L+1)^2}   + O(k)\,.
\end{multline}

We displayed in Table~\ref{Tab_beta3_sigmaL} the asymptotic coefficients $\beta_3$ (in fm) and the zero-energy partial cross sections  provided by Eq.~(\ref{sigma_LSJ})
 (in $\mu$b) for the lowest angular momentum states. They correspond to the Friar+Dipole $n$ form factors.

\bigskip
Our first remark concerns the asymptotic coefficient $\beta_3$. 
 As one can see, the triplet natural parity states ($^3$L$_{J=L}$) have $\beta_3$ independent of $L$, while for 
the unnatural parity states  $^3$L$_{J=L\pm1}$ states, $\beta_3$ increases with $L$.
This $\beta_3$(LSJ=L$\pm$1) increasing   is due to both the tensor and the spin-orbit contributions. The tensor contribution increases with L but converges to a finite value when  $L\to \infty$ since the tensor matrix elements $S_{12} \to -1$.
However the contribution to $\beta_3$(L,S,J=L$\pm$1) due to the spin-orbit term increases linearly with $L$, due the $\lambda_{\pm}(L)$ eigenvalues~(\ref{lambda_pm}).

\bigskip
Our second remark concerns the non-vanishing zero-energy cross section $\sigma_L(0)$. 
They all decrease with increasing $L$ but for the lowest value of $L$ represented in Table  \ref{Tab_beta3_sigmaL}, they are
comparable to  the S-wave partial cross sections described in the previous section and  which have typical values of $0.4-0.2$ $\mu$b. 
This is one of the most striking difference with respect the usual scattering by short-range potentials.

\bigskip
A final remark concern  the contribution to the total zero-energy cross section from the triplet $L>0$ states, as it follows from Eq.~(\ref{sigma_LSJ}),
and that will be written for latter convenience  in the form.
\begin{eqnarray*} \sigma_T (k) &=&\sum_{J=0} \sigma_J \quad {\rm with} \\   \sigma_J &=&   \sigma_{L=J-1,1,J}+ \sigma_{L=J,1,J}+\sigma_{L=J+1,1,J} \,. \end{eqnarray*}  
 
If $\beta_3$ would be independent of $L$, as it is implicitly assumed in {\cite{Gao_PRA59_1999,Muller_PRL110_2013},  the zero-energy cross section $\sigma_L(0)$
would decrease asymptotically as $1/J^3$ for all  states $L=J-1,J,J+1$ and
one could easily obtain the total low-energy cross section. For instance, for natural parity states (L=J) one has
 \[\sigma_T(k)=   {\pi  \beta_3^2\over4} \; \sum_{J=1}^{\infty}  {2J+1\over J^2(J+1)^2}  = {\pi  \beta_3^2\over 4}\,. \]
This is however not the case in the $ln$ system.
In particular, the contribution to the total cross section due  the unnatural parity states, is affected by a  quadratic dependence  on J due to $\beta_3(L=J\pm1)$ and 
according to \eqref{sigma_LSJ} one has
\begin{equation}
 \sum_{J=1}^{\infty} \sigma_J(0)   \sim   \sum_{J=1}^{\infty}    {1  \over J}        
 \end{equation}
which is logarithmically divergent with $J$.
This fact suggests a non integrability of the total differential cross section, and could be either an intrinsic property of the $1/r^3$ potentials with spin-orbit force,
or a consequence of a too restrictive hypothesis in the derivation of (\ref{tgd_LSJ}). References \cite{Gao_PRA59_1999,Muller_PRL110_2013} are indeed
 based on  the Born approximation with the asymptotic $1/r^3$ potentials. It is not clear
 that this approximation could apply when  the asymptotic coefficient $\beta_3$ of these potentials is very large, even linearly diverging with $L/J$.
Work is in progress to clarify this point.

\section{Concluding Remarks}\label{Sect_Conclusion}

We have presented a lepton-neutron potential in configuration space based on the Coulomb interaction
between the pointlike lepton and the neutron charge density plus the hyperfine Hamiltonian
integrated over the neutron electric and magnetic densities.
It is given in the operator form and has a central, spin-spin, tensor and spin-orbit terms, all regulars at the origin and the two latter
displaying a long-range $1/r^3$ tail, precluding the existence of low-energy parameters in non-zero angular momentum states.
Several  parametrisations of the experimentally measured neutron form factors have been used to check
the stability of the predictions.

The S-wave  lepton-neutron low-energy parameters -- coherent and incoherent scattering length and effective range -- have been obtained 
as well as the  corresponding  cross section.
The coherent scattering of $n$ with "electrons-bound-to-atoms" has been considered and the predictions
of the potential have been found in agreement with the experimentally measured value of the coherent n-atom scattering length $b_{ne}=1.23\pm0.03$ fm.
  {To our knowledge, and apart from this latter quantity,  none of the lepton-neutron  low-energy parameters have been already  predicted
and remain experimentally unknown.}

The higher angular momentum states are all coupled in the partial wave $ LSJ$ basis, either by tensor force for the triplets unnatural parity states ($^3$L=J-1$_J-^3$L=J+1$_J$) 
or by spin-orbit term for the single and triplet natural parity states ($^1$L$_{J=L}$ and $^3$L$_{J=L}$ ).
By neglecting this coupling, we have estimated  the low-energy cross section for the lowest partial waves and pointed out 
 a divergence in  the partial wave expansion of the total cross section. The origin of this  behaviour
lies in the spin-orbit interaction for the triplet unnatural parity states, from the combined effect of its long-range tail and the increasing matrix
elements with the angular momentum.

  {The lepton-neutron potentials  presented in this work, which are largely dominated by the magnetic terms (tensor and spin-orbit), can be useful as theoretical inputs in the analysis of 
the precision atomic spectroscopy data with $e$'s and $\mu$'s beyond the H case. 
In particular, to extract  the nuclear charge radii   taking into account the impact  of the neutron electromagnetic structure on the
electron-nucleus interaction.
It is worth mentioning that, contrary to  what happens in H isotopes (proton \cite{p_size_Nature486_2010,muH_Antognini_ARNPS72_2022} and deuterium \cite{mud_CREMA_Science353_2016}),  there is no any
significant difference between the $e$ and $\mu$  results in the $^4$He charge radius  \cite{mu_4He_Nature589_2021}.  
A possible reason for that could be the average of the lepton-nucleon magnetic effects that take place in the $\alpha$-particle but is absent in proton and deuterium.}

\section*{Acknowledgments}
  {The authors are indebted to R. Lazauskas, B. Moussalam, D.R. Phillips and H. Sadjan
 for the very helpful discussions during the elaboration of this work.}
J.C. thanks the  financial support from FAPESP (Funda\c c\~ao de Amparo \`a Pesquisa do Estado de S\~ao Paulo) grant  2022/10580-3. 
T. F. thanks
the financial support from  CNPq (Conselho Nacional de Desenvolvimento Cient\'ifico e Tecnol\'ogico) grant  306834/2022-7,  CAPES (Coordena\c{c}\~ao de Aperfei\c{c}oamento de Pessoal de N\'ivel Superior), Finance Code 001, CAPES/COFECUB  grant 88887.370819/2019-00, and FAPESP (grants 
 2017/05660-0 and 2019/07767-1).  This work is a part of the
project Instituto Nacional de  Ci\^{e}ncia e Tecnologia - F\'{\i}sica
Nuclear e Aplica\c{c}\~{o}es  Proc. No. 464898/2014-5.
We have benefited from the French IN2P3 support to PUMA theory project.

\appendix

\section{Matrix elements of spin-orbit term}\label{Ap_A}

Let us consider the spin-orbit operator in the form
\begin{multline}\label{Lsn}
 \vec{L} \cdot \vec{s}_n= {1\over2}  \left[  j_n(j_n+1) - L(L+1) - {3\over 4} \right]\, , \\ \vec{j}_n= \vec{L}+\vec{s}_n\, ,   \qquad \vec{J}= \vec{j}_n+\vec{s}_e
 \end{multline}
It is diagonal in the eigenbasis
\begin{multline*} \mid L,j_n,J=L,L\pm1\rangle \equiv \\ \mid  [L,s_n=\tfrac12]_{j_n=L\pm\tfrac12}, s_e=\tfrac12 ; J={{j_n}\pm \tfrac12}  = L,L\pm 1 \rangle \, ,  \end{multline*}
the spin-orbit matrix elements are:
\begin{small}
\begin{equation}
\label{Lsn_LjnJBasis}
\hspace{-.2cm} \langle L,j_n=L\pm\tfrac12,J \mid \vec{L} \cdot \vec{s}_n  \mid L, j'_n=L\pm\tfrac12,J\rangle= \delta_{jj'}\lambda_{\pm}(L) 
\end{equation}
\end{small}
for $\forall J =L,L\pm1$,
with eigenvalues given by the right-hand side of (\ref{Lsn}):
\begin{equation}\label{lambda_pm} 
\lambda_{\pm}(L) =   
                  \left\{ \begin{array}{lcl}  {L\over2} &{\rm if} &j_n=L+{1\over2} \cr -{L\over2}-{1\over2} & {\rm if} & j_n=L-{1\over2}\end{array}   \right.
\end{equation}

It is however interesting to know the matrix elements of the spin-orbit operator in the usual partial wave (PW) basis  $\mid SLJ\rangle$:
\begin{multline}\label{Lsn_SLJBasis}
 \langle SLJ \mid   \vec{L} \cdot \vec{s}_n \mid S' L J\rangle=\\ \sum_{j_n=L\pm 1/2}   \langle SLJ \mid Lj_n J\rangle \; \lambda_{\pm}(L)  \; \langle L j_n J \mid S' L J\rangle\,.
\end{multline}
The relation between the $\mid  L, j_n,J \rangle$  and the  $\mid S,L,J\rangle$ basis is given by the 6j coefficients~\cite{Messiah_QM_T2}
\begin{multline}\label{Rel_Basis}
\langle [(s_es_n)_SL]_J \mid [s_e (s_nL)_{j_n}] J\rangle = \\ (-1)^{J+L+1} \; \sqrt{ (2S+1)(2j_n+1)}   \sixj{s_e}{s_n}{S}{L}{J}{j_n}\,  ,
\end{multline}
with $s_e=s_n=\tfrac12$.
According to that, {    most of   $\mid L,j_n,J \rangle$ states corresponds  to a single  $\mid SLJ \rangle$ state}.
Thus for S-wave one has :
\begin{equation}\label{Lsn_SLJBasis_1}
\begin{array}{l l l c  c l}
\mid 0,j_n=\tfrac12,J=0 \rangle                  &\equiv& \mid ^1S_0 \rangle               \cr \cr
\mid 0,j_n=\tfrac12,J=1 \rangle                  &\equiv& \mid ^3S_1 \rangle               
\end{array}
\end{equation}
and for  $L>0$ unnatural parity states 
\begin{equation}\label{Lsn_SLJBasis_2}
\mid L,j_n=L\pm1/2,J=L\pm1 \rangle  \equiv \mid ^3L_{L\pm1}\rangle   \,,   
\end{equation}
where we used  the  spectroscopic notation $$\mid ^{2S+1}L_J\rangle\equiv \mid SLJ \rangle\,.$$

The corresponding matrix elements of the spin-orbit operator are
\begin{eqnarray}
 \langle  ^1S_0           \mid  & \vec{L} \cdot \vec{s}_n   &\mid ^1S_0\rangle                =0  \\
 \langle  ^3S_1            \mid &  \vec{L} \cdot \vec{s}_n & \mid ^3S_1\rangle                = 0   \\
\langle   ^3L_{L\pm1} \mid  & \vec{L} \cdot \vec{s}_n  &\mid ^3L_{L\pm1}\rangle =\lambda_{\pm}(L)  
\end{eqnarray}

It turns out, however, that {    the two natural parity states,  $\mid L,j_n=L\pm\tfrac12,J=L\rangle$, 
are (orthogonal) linear combinations of the corresponding singlet and triplet natural parity states}, i.e. $\mid S=0,L,J=L \rangle$ and $\mid S=1,L,J=L \rangle$:
\begin{multline}   \begin{pmatrix}  \mid L,j_n=L-\tfrac12, J=L\rangle\cr  \mid L, j_n=L+\tfrac12,J=L\rangle \end{pmatrix}= M_L\begin{pmatrix}  \mid {^1}L_L\rangle\cr  \mid  {^3}L_L\rangle  \end{pmatrix}\,, \\ \\ 
 M_L= \begin{pmatrix} -\sqrt{L \over 2L+1} & \sqrt{L +1\over 2L+1} \cr \sqrt{L +1\over 2L+1}  & \sqrt{L \over 2L+1} \end{pmatrix}\,,   \end{multline}
with, det($M_L$)=-1 and  $M_L^{-1}=M_L$. 

The matrix elements of the spin-orbit operator in this basis are obtained by inserting  relation  (\ref{Rel_Basis}) into (\ref{Lsn_SLJBasis}) and read:
\begin{small}
\begin{equation}\label{Lsn_SLJBasis_3}
\hspace{-.3cm} \langle ^{2S+1}L_{J=L} \mid   \vec{L} \cdot \vec{s}_n \mid ^{2S'+1}L_{J=L}\rangle
 =  M_L  \begin{pmatrix}  \lambda_- & 0 \cr 0 & \lambda_+ \end{pmatrix}  M_L \; ,
\end{equation}
\end{small}
which leads to Eq.~\eqref{Lsn_SLJ_NPS}.
As one can see, the singlet  $\mid ^1$L$_J\rangle$ and the triplet  $\mid ^3$L$_J\rangle$ states with natural parity (i.e. with $J$=$L$) are coupled by the spin-orbit term (\ref{Lsn}),
which generates transitions between them. This is the case for $^1$P$_1-^3$P$_1$, $^1$D$_2-^3$D$_2$,$^1$F$_3-^3$F$_3$ etc.

\end{document}